\documentclass[aps,pra,twocolumn,superscriptaddress,a4paper,english]{revtex4-1}

\usepackage{times}
\usepackage{color}
\usepackage[colorlinks=true,urlcolor=blue,citecolor=blue]{hyperref}
\usepackage{url}
\usepackage{breakurl}
\usepackage{soul}
\usepackage{graphicx}
\usepackage{amsmath}
\usepackage{subfigure}
\usepackage{xcolor}
\usepackage{amssymb}
\usepackage{latexsym}
\usepackage{hyperref}
\newcommand {\R}{\textcolor {black}}

\begin{document}

\title{Efficient Quantum Simulation for Thermodynamics of Infinite-size Many-body Systems in Arbitrary Dimensions}

\author{Shi-Ju Ran}
\email[Corresponding author. Email: ]{sjran@cnu.edu.cn}
\affiliation{Department of Physics, Capital Normal University, Beijing 100048, China}
\affiliation{ICFO - Institut de Ciencies Fotoniques, The Barcelona Institute of Science and Technology, Av. Carl Friedrich Gauss 3, 08860 Castelldefels (Barcelona), Spain}

\author{Bin Xi}
\affiliation{College of Physics Science and Technology, Yangzhou University, Yangzhou 225002, China}

\author{Cheng Peng}
\affiliation{School of Physical Sciences, University of Chinese Academy of Sciences, P. O. Box 4588, Beijing 100049, China}

\author{Gang Su}
\affiliation{School of Physical Sciences, University of Chinese Academy of Sciences, P. O. Box 4588, Beijing 100049, China}
\affiliation{Kavli Institute for Theoretical Sciences and CAS Center for Excellence in Topological Quantum Computation}

\author{Maciej Lewenstein}
\affiliation{ICFO - Institut de Ciencies Fotoniques, The Barcelona Institute of Science and Technology, Av. Carl Friedrich Gauss 3, 08860 Castelldefels (Barcelona), Spain}
\affiliation{ICREA, Pg. Llu\'is Companys 23, 08010 Barcelona, Spain}


\date{\today}




\begin{abstract}
In this work we propose to simulate many-body thermodynamics of infinite-size quantum lattice models in one, two, and three dimensions, in terms of few-body models of only O(10) sites, which we coin as quantum entanglement simulators (QES's). The QES is described by a temperature-independent Hamiltonian, with the boundary interactions optimized by the tensor network methods to mimic the entanglement between the bulk and environment in a finite-size canonical ensemble. The reduced density matrix of the physical bulk then gives that of the infinite-size canonical ensemble under interest. We show that the QES can, for instance, accurately simulate varieties of many-body phenomena, including finite-temperature crossover and algebraic excitations of the one-dimensional spin liquid, the phase transitions and low-temperature physics of the two- and three-dimensional antiferromagnets, and the crossovers of the two-dimensional topological system. Our work provides an efficient way to explore the thermodynamics of intractable quantum many-body systems with easily accessible systems.
\end{abstract}
\maketitle

\section{Introduction}

Simulating quantum many-body physics is one of the central tasks in condensed matter physics and the related fields such as quantum information/quantum simulation. While the analytical solutions are extremely rare, one important method to study  many-body systems consists in classical simulations. Paradigm approaches include the exact diagonalization (cf. \cite{raventos2017cold}), quantum Monte Carlo (QMC) \cite{CA86QMCrev,FMNR01QMCrev}, density matrix renormalization group (DMRG) \cite{W92DMRG,W93DMRG}, and tensor network (TN) methods \cite{VMC08MPSPEPSRev,CV09TNSRev, S11DMRGRev,O14TNSRev,HV17TMTNrev,RTPC+17TNrev}. However, due to the high complexity of the quantum many body problems, there are still many unsettled issues that cannot be reliably accessed by classical simulations. For instance, the possible candidate models for quantum spin liquids \cite{M00QSLrev,B10QSLRev,W12QSL,SB17QSLRev} in two and higher dimensions, as well as  efficient algorithms to simulate their thermodynamics \cite{O12CTMRG, RLXZS12ODTNS, CCD12FTPEPS, RXLS13NCD, CD15TPO, CD15FTPEPS, CDO16TPO, CRD16TPO, CDO17TPOQMC} are still under very hot debate.

Even for the many-body systems that are theoretically well understood, it is still a challenge to realize them in a controllable manner, and  to demonstrate certain targeted many-body features in experiments. A common way to do it is to search for materials occurring naturally. To name a few, the compounds $\alpha$-RuCl$_3$ \cite{BBYA+15QSLexp}, YbMgGaO$_4$\cite{LSWL+16QSLexp} and ZnCu$_3$(OH)$_6$FBr \cite{HHCNR+12fractionalized} are the rare examples of spin-liquid candidates, which can be described by Kitaev model on honeycomb lattice, and Heisenberg model on triangular and kagom\'e lattices, respectively. Without natural materials, it is extremely difficult to realize the targeted many-body features in experiments by purposefully designing microscopic interactions in a material.

A much more flexible approach is to use quantum simulators (c.f. \cite{LSA12SimuBook,CZ12QsimuRev, BDN12Qsimu, AW12simu, GAN14simu, ABBCE+18QTroadmap}), which are defined as the controllable and simple quantum systems that mimic quantum models of high complexity. This approach allows to design different interactions with the same experimental platform such as cold, or ultra-cold atoms/ions. Various phenomena have been successfully simulated by quantum simulators, including  Bose-Einstein condensation (e.g., \cite{DMAD95BEC,anderson1995observation, GWHSP05BEC}), quantum magnets (e.g., \cite{FSGPS08qsimuexp, SBMT+11AFsimu}), strongly correlated electrons (e.g., \cite{JSGME08qsimu, CJMGG18fermionSimu}), and so on. Quantum simulators offer a promising way two realize the targeted physics in controllable systems. One bottleneck for simulating lattice models is the difficulty of reaching large sizes, which hinders the exploration of the physics in the thermodynamic limit.

\begin{figure}
	\includegraphics[angle=0,width=1\linewidth]{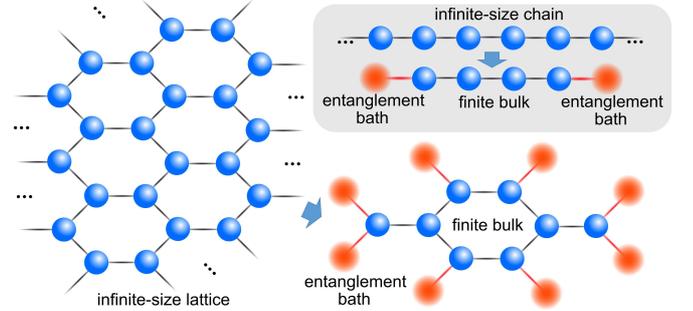}
	\caption{(Color online) Illustrations of the quantum entanglement simulators of 1D (upper-right inset) and 2D systems.}
	\label{fig1}
\end{figure}

In this work we show that with the few-body systems of only O(10) sites, coined as quantum entanglement simulators (QES's), the thermodynamics of infinite-size many-body systems in one and two dimensions can be accurately simulated. The key idea is to introduce the entanglement-bath sites on the boundary, and optimize their interactions with the physical sites \cite{RPPSL17AOP3D} (Fig. \ref{fig1}). A QES is described by a simple Hamiltonian that reads
\begin{eqnarray}
\hat{H}_{\text{QES}} = \sum_{\langle i,j \rangle \in \text{bulk}} \hat{H}^{[i,j]} + \sum_{\langle i \in \text{bulk},n \in \text{bath} \rangle } \hat{\mathcal{H}}^{[i,n]}.
\label{eq-hamilt}
\end{eqnarray}
$\hat{H}^{[i,j]}$ denotes the two-body Hamiltonian on the $i$-th and $j$-th physical sites in the bulk, and $\hat{\mathcal{H}}^{[i,n]}$ denotes the two-body Hamiltonian between the $i$-th physical and $n$-th bath sites. Here, we restrain ourselves to the nearest-neighboring interactions.

The aim is to mimic the thermodynamics of the infinite-size model with the Hamiltonian
\begin{eqnarray}
\hat{H} = \sum_{\langle i,j \rangle} \hat{H}^{[i,j]}.
\label{eq-hamiltInf}
\end{eqnarray}
The physical information is extracted from the reduced density matrix
\begin{eqnarray}
\hat{\rho}_R = \text{Tr}_{\text{bath}} \hat{\rho},
\label{eq-rho}
\end{eqnarray}
with $\text{Tr}_{\text{bath}}$ the trace over the degrees of freedom of the bath sites. $\hat{\rho}$ is the density matrix of the QES. We have $\hat{\rho} = |\Psi \rangle \langle \Psi|$ for the ground-state simulation (with $|\Psi \rangle$ the ground state of $\mathcal{\hat{H}}$) and $\hat{\rho} =e^{-\beta \mathcal{\hat{H}}}$ for the simulation at the inverse temperature $\beta$. $\hat{\rho}_R$ mimics the reduced density matrix of infinite-size system that traces over everything except the bulk. 

On the boundary, $\{\hat{\mathcal{H}}^{[i,n]}\}$ are optimized from the ground state to mimic the entanglement between the finite-size bulk and the rest in the infinite-size system \cite{RPPSL17AOP3D}. $\{\mathcal{\hat{H}}^{[i,n]}\}$ are optimized by the infinite DMRG algorithm \cite{W92DMRG,W93DMRG} for 1D cases, or its variants on infinite-size tree TN \cite{LCP00TreeDMRG,NC13TTN, RPPSL17AOP3D} for 2D and 3D cases \footnote{The codes for obtaining the physical-bath Hamiltonians $\hat{\mathcal{H}}$ with the necessary instructions can be found at \url{https://github.com/ranshiju/FT-QES}.}. The dimension $D$ of the entanglement bath controls the total number of eigenstates, or in other words, the upper bound of the entanglement between the bulk and environment. 

After obtaining $\hat{H}_{\text{QES}}$, the simulation of the infinite-size model becomes that of the finite-size one. Since the size of $\hat{H}_{\text{QES}}$ is small and the interactions are only two-body nearest neighbors, one way to simulate $\hat{H}_{\text{QES}}$ is to use techniques such as cold atoms to do quantum simulations. Here, we generalize the linearized tensor renormalization group (LTRG) algorithm \cite{LRGZXY+11LTRG} with sufficiently large bond dimensions ($\chi = 400 \sim 600$) to numerically simulate the $\hat{H}_{\text{QES}}$.

The performance of the QESs is testified on one- and two-dimensional quantum lattice models, i.e, the XY chain and the Heisenberg model on honeycomb lattice. The accuracy of the QES's is shown to be one or two orders of magnitude higher than the models of the same size, but without the entanglement bath. We show that the QES can accurately mimic the finite-temperature cross-over and the low-temperature algebraic excitations of 1D spin liquid, and the N\'eel-paramagnetic phase transition and low-temperature gapped excitations of 2D anti-ferromagnet. {\R{We further apply the QES for the 3D Heisenberg model on cubic lattice.}} Our work provides a theoretical scheme of modeling novel quantum simulators that can accurately reproduce the targeted many-body features in the thermodynamic limit by a small number of sites. The QESs can be considered as prototype models to (theoretically or experimentally) simulate strongly correlated systems of infinite size.

\section{Benchmark results}

\begin{figure}
	\includegraphics[angle=0,width=1\linewidth]{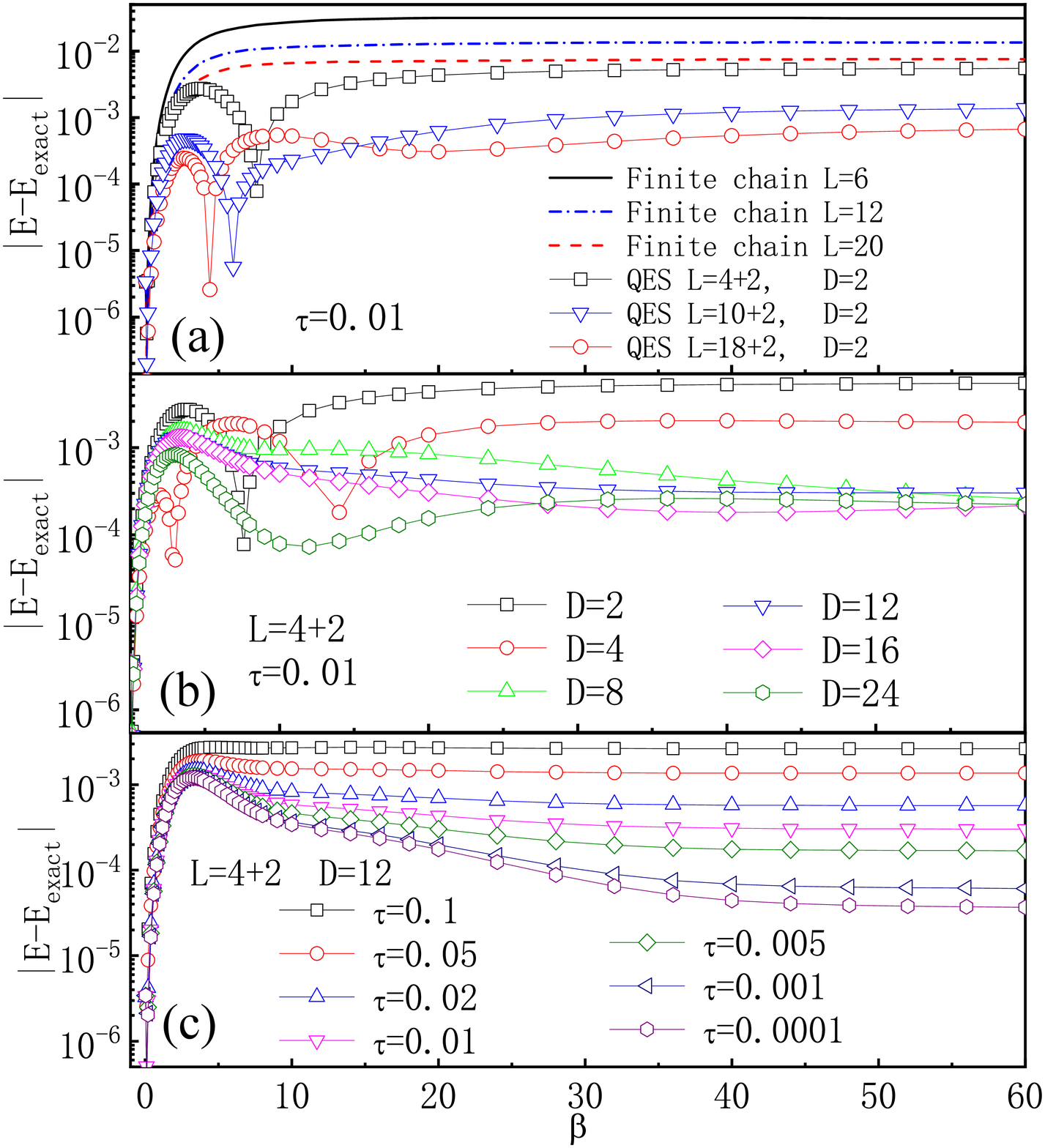}
	\caption{(Color online) The errors of the QES's that simulate the infinite-size 1D XY chain, comparing with the exact solution and the finite-size models without entanglement bath. The error versus the inverse temperature $\beta$ for different (a) sizes $L$, (b) entanglement-bath dimensions $D$, and (c) Trotter steps $\tau$ are testified.}
	\label{fig2}
\end{figure}

We first simulate the infinite-size XY chain as a benchmark, where the Hamiltonian reads $\hat{H} = \sum_{i} (\hat{S}^{[i]}_x \hat{S}^{[i+1]}_x + \hat{S}^{[i]}_y \hat{S}^{[i+1]}_y)$ with $\hat{S}^{[i]}_{\alpha}$ is the $\alpha$ component of the spin-$1/2$ operators ($\alpha=x$, $y$, $z$). We take the Planck and Boltzmann constants $\hbar = k_B = 1$ for convenience. Its thermodynamics can be analytically solved \cite{LSM61exact}. Fig. \ref{fig2} (a) shows the error of energy per site $|E-E_{\text{exact}}|$ at different inverse temperature $\beta$ (note $E = \text{Tr}(\hat{\rho}_R \sum_{\langle i,j \rangle \in \text{bulk}} \hat{H}^{[i,j]})/(N \text{Tr} \hat{\rho}_R)$ with $N$ the number of the physical sites). We take the bath dimension $D=2$ so that the bath sites are effectively spin-$1/2$'s. With the same size, the accuracy of the QES is about one order of magnitude higher than that of the normal finite-size chain. In other words, the QESs with 6 spins performs even slightly better than the normal chain with 20 spins. Note that the dips of the curves are due to the change of the signs of $E-E_{\text{exact}}$.

The accuracy can be improved by increasing the bath dimension $D$ [Fig. \ref{fig2} (b)]. We take four physical sites as the bulk and two bath sites on the boundary. The reason is that the QES with a higher $D$ can capture more correlations and entanglement. Using the time matrix product state, it is known that $\xi \propto D^{c}$ with $\xi$ the upper bound of the dynamic correlation length the QES can capture and $c$ a state-dependent positive constant \cite{TOIL08EntScaling, TTLR18tMPS}. Therefore, a finite $D$ might lead to the loss of the long-range correlations, which we call the \textit{bath correlation error}.

For $D \geq 8$, the error converges to be around O($10^{-4}$). This is in fact the \textit{Trotter error} [$\sim$O($\tau^2$)] \cite{SI87Trotter, IS88Trotter}, as in LTRG we take the finite-temperature density operator as $\hat{\rho}(\beta) = [\prod \exp(-\tau \hat{H}^{[i,j]}) \prod \exp(-\tau \hat{\mathcal{H}}^{[i,n]})]^K + O(\tau^2)$ with $\beta = K\tau$. Unlike the bath correlation error that is the error of the QES itself, Trotter error is a computational error of the LTRG algorithm that appears in the numerical simulations of the QES's. Fig. \ref{fig2} (c) shows the error of different $\tau$'s. Note that the computational cost increases linearly with $1/\tau$. The results are consistent with the fact that the Trotter error accumulates as $\beta$ increases and finally converges to the Trotter error at zero temperature (ground state).

There exists one more error related to the dynamic correlation length. In a QES, we fixed the physical-bath Hamiltonians $\hat{\mathcal{H}}^{[i,n]}$ to be the same for all temperatures, which is optimized from the ground-state calculation. This is a reasonable assumption so that the QES is described by a well-defined low-energy effective Hamiltonian. If the dynamic correlation length $\xi$ is much shorter than the inverse temperature $\beta$, $\hat{\mathcal{H}}^{[i,n]}$ should indeed be temperature-independent. But $\hat{\mathcal{H}}^{[i,n]}$ might depend on $\beta$ when $\beta$ is comparable to $\xi$. It is responsible to the peak of the error near the crossover point (see Fig. \ref{fig2}). One way to improve the accuracy is to optimize $\hat{\mathcal{H}}^{[i,n]}$ according to the targeted temperature by using, e.g., the imaginary-time sweep algorithm \cite{RXLS13NCD}. We opt not to do so because the QES will not have a well-defined Hamiltonian if $\hat{\mathcal{H}}^{[i,n]}$ is temperature-dependent. We dub the error from the temperature independence as \textit{thermal correlation error}. The bath and thermal correlation errors are together coined as the correlation errors.

\begin{figure}
	\includegraphics[angle=0,width=1\linewidth]{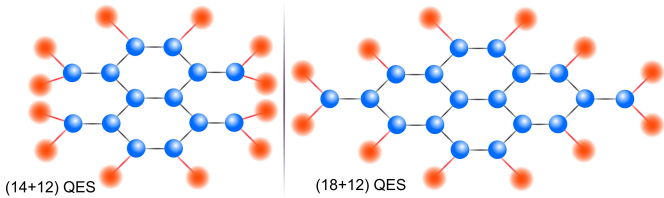}
	\caption{(Color online) The (14+12) and (18+12) QES's for the Heisenberg model on honeycomb lattice.}
	\label{fig3}
\end{figure}

For two-dimensional systems, we simulate the Heisenberg model on infinite-size honeycomb lattice, where the Hamiltonian reads $\hat{H} = \sum_{\langle i,j \rangle} (\hat{S}^{[i]}_x \hat{S}^{[j]}_x + \hat{S}^{[i]}_y \hat{S}^{[j]}_y + \hat{S}^{[i]}_z \hat{S}^{[j]}_z)$. We construct three QES's of different numbers of physical and bath sites, dubbed as (8+8), (14+12), and (18+12) QES's (Figs. \ref{fig1} and \ref{fig3}).

\begin{figure}
	\includegraphics[angle=0,width=1\linewidth]{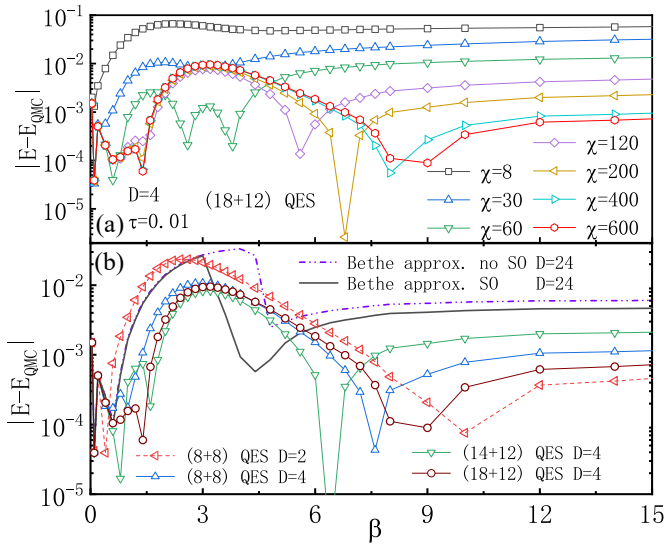}
	\caption{(Color online) The energy discrepancy of the Heisenberg model on honeycomb lattice by comparing with QMC results (extrapolated to infinite size). In (a), we show the discrepancy of the (18+12) QES ($D=4$) with different dimension cut-offs $\chi$ of the LTRG. (b) We present the results of (8+8), (14+12) and (18+12) QES's with $D=2$ and $4$. The results by the Bethe approximations (with and without super-orthogonalization) are also presented for comparison. We take $\tau=0.01$ and $\chi \geq 400$ for the LTRG simulations.}
	\label{fig4}
\end{figure}

Fig. \ref{fig4} shows the discrepancy of the energy per site between the results of the QES's by LTRG and QMC. In this work, all the QMC results are obtained by extrapolating the size to infinity. Fig. \ref{fig4} (a) shows the discrepancy with different bond dimensions $\chi$ of LTRG. Our results imply that the computational error of LTRG due to the finiteness of $\chi$, called the \textit{truncation error}, is much more significant than the 1D cases. By increasing $\chi$ to the maximum that our computational resources can tolerate, the energy discrepancy is O($10^{-2}$) near the crossover point, and O($10^{-3}$) at the low temperatures. Note that the QES, if realized by quantum simulation or quantum computation, will not suffer the computational errors, and the accuracy should be much more higher than our numerical results.

Besides the errors of the QES's of 1D chains, there exists the \textit{structure error} for 2D systems that can be understood as the following. Considering that the coordination number of honeycomb lattice is $z=3$, one can define a super lattice called Bethe lattice \cite{B1935Bethe} also with $z=3$. If one checks only locally (e.g., one site and its neighbors), there will be no difference between these two lattices. The differences appear when one meets the loops in the honeycomb lattice, as there are no loopy structures in the Bethe lattice. This means that the model defined on the $z=3$ Bethe lattice is a zero-loop approximation of that on honeycomb lattice (sometimes known as the Bethe approximation, or simple update scheme in the sense of the TN-states variational methods \cite{JWX08SimpleUpdate}), where the error is due to the destructions of loopy structures \cite{RXLS13NCD}. For the Bethe lattice, the entanglement bath sites as well as the physical-bath interactions $\hat{\mathcal{H}}^{[i,n]}$'s can be efficiently calculated by the renormalization group (RG) flows using a generalized DMRG algorithm \cite{LCP00TreeDMRG,NC13TTN}. The dimension of the RG flow is the dimension of the entanglement bath site $D$ \cite{R16AOP,RPPSL17AOP3D}. Such $\hat{\mathcal{H}}^{[i,n]}$'s are put on the boundary of a QES, meaning that the environment is approximated by Bethe lattice, suffering from the structure error. Within the bulk, all interactions are fully considered, implying there is no structure error inside. This is why the QES possesses a higher accuracy than the Bethe approximation.

The results shown in Fig. \ref{fig4} (b) support the above analyses. The Bethe approximations even with $D=24$ show the highest discrepancy in our simulations. By using the super-orthogonalization (SO) trick \cite{RLXZS12ODTNS}, the optimal point of the Bethe approximation is better reached and the discrepancy is slightly lowered. For the (8+8) QES only with $D=2$, the discrepancy near the crossover temperature is comparable with the Bethe approximation of $D=24$ and is much smaller as $\beta$ increases. It indicates the structure error is lowered due to the loopy structure inside the bulk of the (8+8) QES.

When we increase the bath dimension to $D=4$ for the (8+8) QES, the discrepancy near the crossover temperature is considerably lowered. This is due to the decrease of the bath correlation error (since other errors should not change by increasing $D$). The discrepancy at lower temperatures becomes larger. This should be due to the increase of the truncation error of LTRG. When we increase the size to (14+12) and (18+12), more loops are contained in the bulk (see Fig. \ref{fig3}), but the discrepancy near the crossover temperature only changes slightly. This implies that the dominant error is no longer structure error but the thermal correlation error. At lower temperatures, the thermal correlation error rapidly decreases as discussed above, where the structure and truncation errors dominate. Note that at lower temperatures, the $D=2$ (8+8) QES seems to give the lowest discrepancy. This should be due to the canceling of the errors of different signs, which is not controllable.

Without the $\hat{\mathcal{H}}^{[i,n]}$'s for comparison, the finite-size and boundary effects in 2D are much more severe than those in 1D. If we take the finite lattices with the same sizes of the QES's, the discrepancy increases with $\beta$ and converges to about $0.16$ (8+8 sites) and $0.11$ (18+12 sites), which are one or two orders of magnitude higher than the QES's. To reach small discrepancies, the size has to be very large. For instance, the discrepancy will reach O($10^{-4}$) with O($10^3$) sites.

\begin{figure}
	\includegraphics[angle=0,width=1\linewidth]{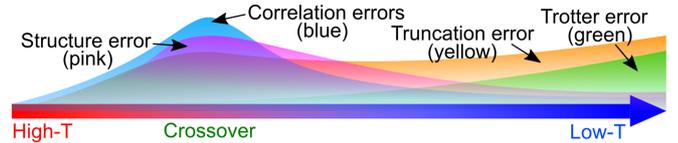}
	\caption{(Color online) The {\R{schematic}} illustration of the errors of a QES, which are (bath and thermal) correlation errors, structure error, Trotter error and truncation error. The errors with a crossover or transition should be similar. {\R{We assume both the high- and low-temperature phases to be gapped.}} See the main text for details.}
	\label{fig5}
\end{figure}

The temperature dependence of the errors is illustrated in Fig. \ref{fig5}. {\R{As the errors of TN algorithms (particularly in higher dimensions) usually cannot be exactly given, this figure is only to schematically indicate the errors.}} At very high temperatures, the system is less correlated or entangled, thus there exit almost no errors except for certain Trotter error. As the temperature decreases, the system gains more and more dynamic correlations and entanglement. Then there appears the thermal correlation error. While approaching the crossover{\R{/critical}} temperature, the correlation errors dominate because the dynamic correlation length reaches the maximum in this region and becomes comparable to {\R{or larger than}} the inverse temperature. At the low temperatures, the thermal correlation error decays rapidly as $\beta$ is much larger than the dynamic correlation length. {\R{Here we assume that the low-temperature phase is gapped.}} Trotter error accumulates as the temperature decreases and may dominate depending on the value of $\tau$.

Another error that may dominate near and below the crossover temperature is the truncation error. For 1D simulations, we take $\chi = 400 \sim 600$, where the results converge with $\chi$ and the truncation error is ignorable. For 2D simulations, the required $\chi$ increases with the bath dimension $D$ and the size of the QES. The truncation error is more severe than the 1D simulations and dominates at finite and low temperatures with the maximal $\chi$ that can be reached by our computers.

The structure error (only for 2D and 3D cases) reaches the maximum near the crossover/critical point. This is because the loops contribute more and more to the physics of the system when longer (spatial) correlations appear. Deep in the low-temperature (gapped) phases, the structure error still exists, but is sub-leading compared with the truncation or Trotter errors.

We shall stress that the truncation and Trotter errors are the computational errors of the LTRG algorithm, not from the QES itself. While our numerical simulations are accurate and reliable, the results can be further improved if one uses better algorithms or quantum simulations that do not suffer from the computational errors.

\section{Simulating many-body thermodynamics}

\begin{figure}
	\includegraphics[angle=0,width=1\linewidth]{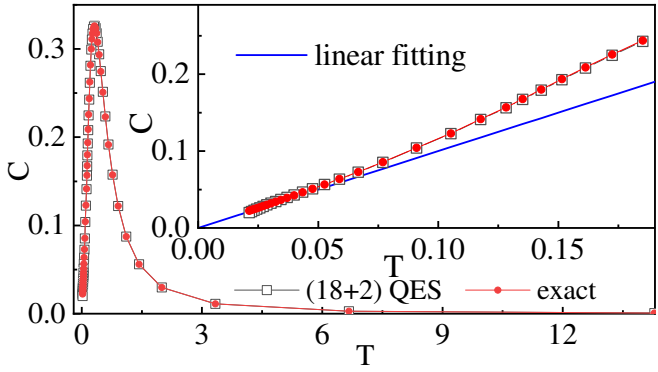}
	\caption{(Color online) The specific heat $C$ versus temperature $T$ of the XY chain obtained by the (18+2) QES and the exact solution. The inset shows the linear scaling behavior at low temperatures due to the gapless excitations.}
	\label{fig6}
\end{figure}

\textbf{Simulating 1D spin liquid}. We now demonstrate that non-trivial many-body physics can be accurately reproduced by small-size QES's. We firstly simulate the XY chain, where the ground state is gapless Tomonaga–Luttinger liquid \cite{H80Luttinger, H81Luttinger}. Fig. \ref{fig6} shows the specific heat $C = \partial E / \partial T$ simulated by a (18+2) QES. At all temperatures, the results from the QES and exact solution coincides remarkably well. The crossover between the high-temperature paramagnetic phase and low-temperature liquid phase is accurately mimicked.

At low temperatures, the QES exhibits linear scaling property, i.e., $C(T) \sim T$ (inset of Fig. \ref{fig6}), which is exactly the property of the low-lying excitations of the gapless Tomonaga–Luttinger liquid. Our simulation shows that such a linear behavior can be captured by the QES down to at least $T \sim 10^{-2}$.

\begin{figure}
	\includegraphics[angle=0,width=1\linewidth]{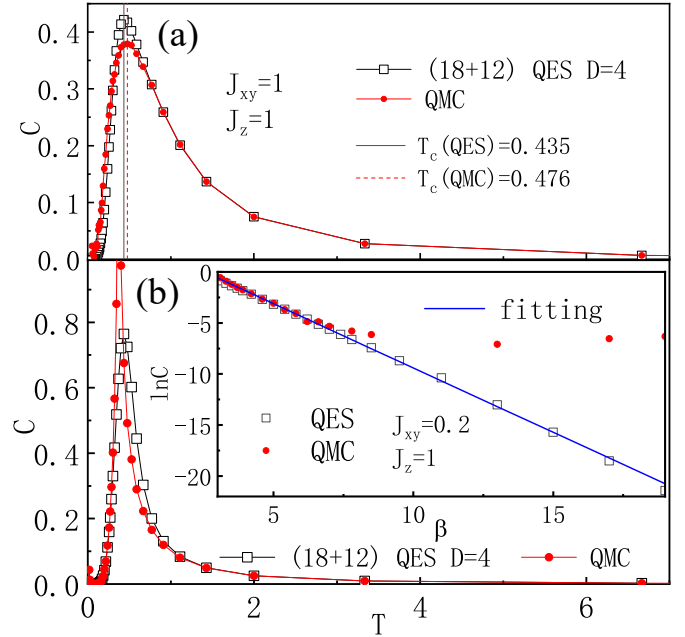}
	\caption{(Color online) The specific heat $C$ versus temperature $T$ of (a) the isotropic Heisenberg model and (b) the XXZ model on honeycomb lattice, calculated by the (18+12) QES and the QMC. The inset of (b) demonstrates the exponential scaling behavior of $C$ against inverse temperature $\beta$ due to the gapped excitations of the antiferromagnetic phase.}
	\label{fig7}
\end{figure}

\textbf{Simulating 2D crossover and phase transition of quantum antiferromagnets}. We simulate the specific heat of the isotropic Heisenberg model on honeycomb lattice. It is well known that the ground state is the gapless N\'eel phase, which is separated from the high-temperature paramagnetic phase by a crossover. Fig. \ref{fig7} (a) shows that the specific heat is accurately given by the QES, compared with the QMC results. The largest discrepancy appears at the crossover point. The crossover temperature is accurately addressed by the QES, with a difference $\sim 10^{-2}$ compared with the QMC results.

We also simulate the XXZ model on honeycomb lattice with the Hamiltonian $\hat{H} = \sum_{\langle i,j \rangle} (J_{x}\hat{S}^{[i]}_x \hat{S}^{[j]}_x + J_{y} \hat{S}^{[i]}_y \hat{S}^{[j]}_y + J_{z} \hat{S}^{[i]}_z \hat{S}^{[j]}_z)$. We take $J_{x}=J_y=J_{xy}=0.2$ and $J_z=1$. At low temperatures, the model spontaneously breaks the Z$_2$ symmetry due to the spin anisotropy and enters the antiferromagnetic phase through an Ising-type phase transition.

Fig. \ref{fig7} (b) shows that such a phase transition is accurately captured by the divergent peak of the specific heat. The discrepancy of the critical point compared with the QCM result is $\sim 10^{-2}$. At low temperatures, the QES faithfully captures the gapped excitations [see semi-log plot in the inset of Fig. \ref{fig7} (b)], where its specific heat exhibits an exponential scaling behavior down to the temperatures $\sim 10^{-2}$. Note that in this region, the worm QMC suffers certain fluctuations and fails to produce the exponential scaling behavior of the specific heat.

\textbf{Simulating crossovers of Kitaev model}.
We show that the QES with a handful of sites can reproduce the properties of Kitaev model \cite{kitaev2006anyons} at finite temperatures. Kitaev model is a well-known model whose ground state possess non-trivial topological orders. The Hamiltonian is written as $\hat{H} = \sum_{\langle ij \in \alpha \rangle} J_{\alpha} \hat{S}^{[i]}_{\alpha} \hat{S}^{[j]}_{\alpha}$ with $\alpha = x, y, z$. The work by Nasu \textit{et al} \cite{NUM15kitaev} shows that at finite temperatures, the system undergoes two crossovers at two temperature scales; the high-temperature crossover is driven by itinerant Majorana fermions, and the low-temperature one is induced by the thermal fluctuation of the fluxes of localized Majorana fermions. 

\begin{figure}
	\includegraphics[angle=0,width=1\linewidth]{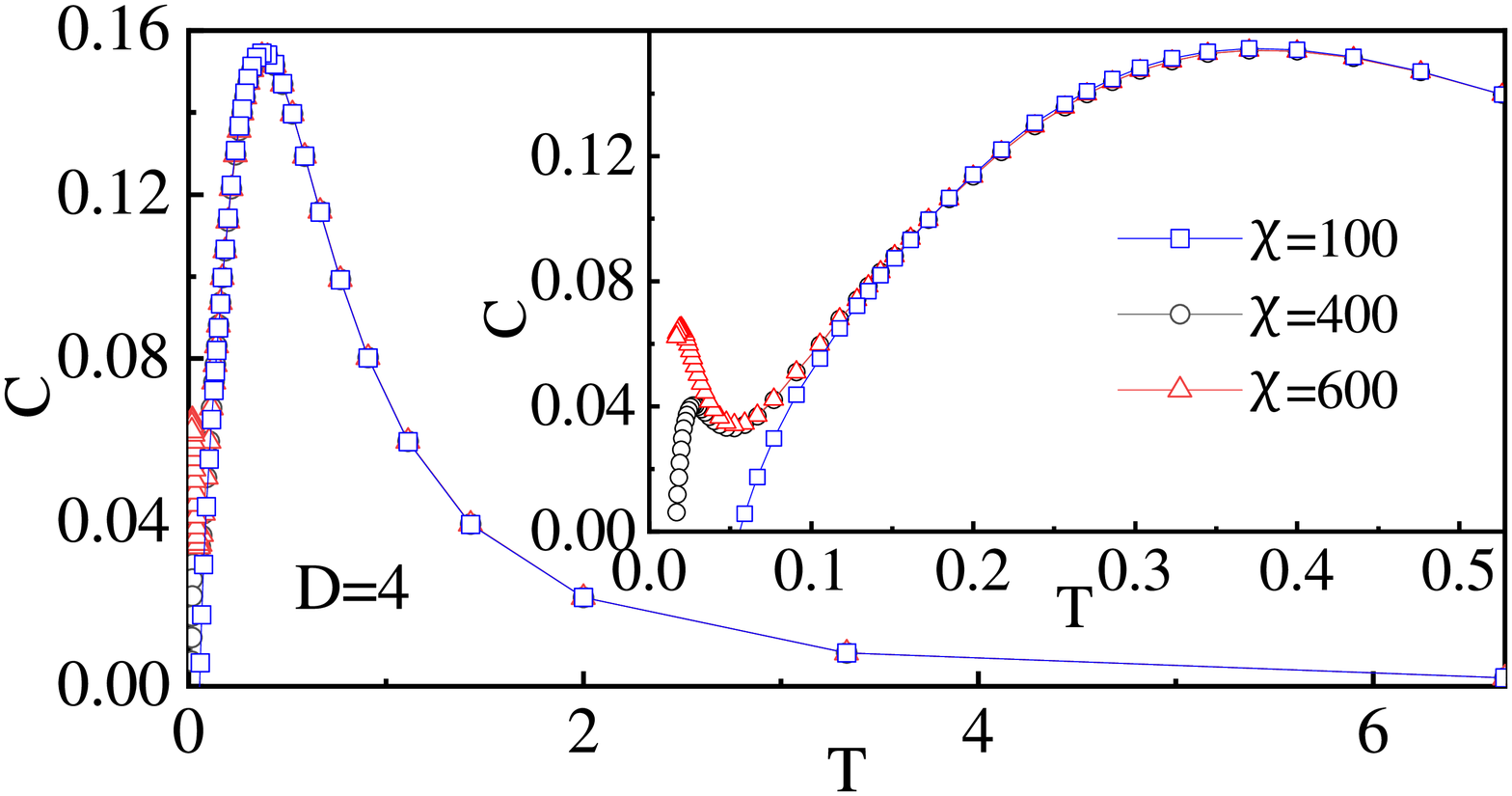}
	\caption{(Color online) The specific heat $C$ of Kitaev model on honeycomb lattice simulated by (18+12) QES with $D=4$. The inset shows zooms in the low-temperature area. The two temperature scales are exhibited by the QES, where it requires much larger bond dimension in LTRG ($\chi>400$ approximately) to capture the low-temperature scale.}
	\label{fig8}
\end{figure}

Fig. \ref{fig8} shows the specific heat of the (18+12) QES with $D=4$ at the isotropic point ($J_x = J_y = J_z$). Our best LTRG simulation ($\chi=600$, $\tau=10^{-2}$) shows that the QES well captures the expected two energy scales. The high-temperature crossover at $T \sim 0.4$ crossover is revealed by LTRG even with small $\chi$'s. The low-temperature crossover is shown at $T \sim 0.02$, which is more challenging to access by LTRG; it requires much larger $\chi$ or the computational errors will be too large to see this crossover from the specific heat. We shall stress that the computational errors are just from classically simulating the QES (e.g., the errors due to the insufficiently large $\chi$ in LTRG), which do not belong to the errors of the QES model. 

\textbf{Simulating 3D thermodynamic phase transition and low-temperature physics}.
For the many-body algorithms such as TN and QMC, 3D quantum systems are obviously more challenging to simulate than 2D. This is because the computational complexity of 3D simulations normally scales much faster than that of 2D systems. {\R{Particularly for the 3D infinite-size systems, the TN simulations are rare even for the ground states \cite{RPPSL17AOP3D, JO18graphPEPS}, and there is currently no efficient TN algorithm reported for the finite-temperature simulations}} to the best of our knowledge.

In Fig 9, we demonstrate the specific heat of the Heisenberg model on cubic lattice, simulated by (8+24) QES with $D=4$. Other sizes can be chosen, for example, $(L^3 + 6L^2)$ sites with $L$ the length of a side of the physical bulk. Consistent with existing results \cite{S98CubicQMC} and our QMC simulations ($T_c \simeq 1.1$), the transition temperature is indicated by a round peak at about $T_c \simeq 0.95$ by the QES's even with small $\chi$'s. The discrepancy of the specific heat itself is quite large, however, since the QES is not large enough to properly capture the long-range correlations near the transition point. To improve the accuracy, one way is to enlarge the bulk of the QES. Still, we shall note that the QES is expected to perform significantly better than the systems of the same size with the (ordinary) open boundary condition. The reason is that the entanglement bath mimics the infinite-size tree-like environment, and the error of QES is mainly from the finite-loop effects instead of the finite-size effects in the ordinary open-boundary systems. At low temperatures where the system is in a gapped phase, the exponential decay of the specific heat is accurately reproduced by the QES; the precision increases with $\chi$. Our best LTRG simulation with $\chi=400$ shows that the exponential scaling of the QES can at least persist down to $\beta=6$, while the QMC simulation losses the exponential behavior for about $\beta>3$.

\begin{figure}
	\includegraphics[angle=0,width=1\linewidth]{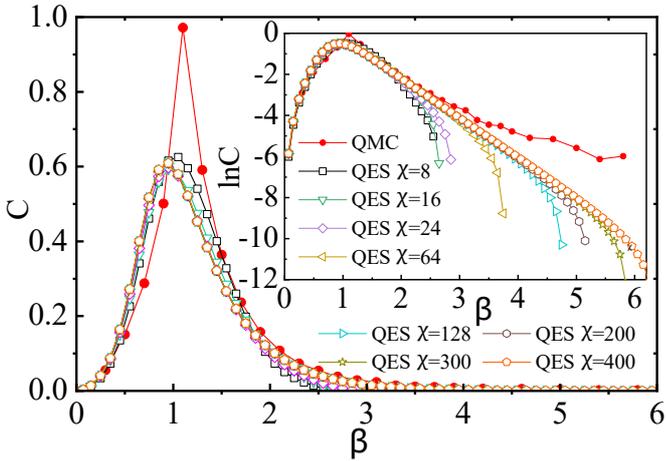}
	\caption{(Color online) The specific heat $C$ of the Heisenberg model on cubic lattice, calculated by QMC and (8+24) QES with different dimension cut-offs $\chi$ in LTRG. We take the bath dimension $D=4$. {\R{The transition temperature by QMC is $T_c \simeq 1.1$ and that by QES is $T_c \simeq 0.95$. The inset shows the semi-log plot.}} The exponential decay of $C \sim e^{-\Delta \beta}$ (with $\Delta$ the gap) at low temperatures is also captured by the QES down to $\beta \simeq 6$ with $\chi = 400$. For comparison, QMC suffers visible instability for about $\beta > 3$. The dominant error of the QES is only the computational errors of LTRG due to the finiteness of $\chi$.}
	\label{fig9}
\end{figure}

\section{Methods}

The construction and optimization of the physical-bath interactions $\hat{\mathcal{H}}^{[i,n]}$'s for the QES were originally proposed in Refs. \cite{R16AOP,RPPSL17AOP3D} to simulate ground states. This work extends the QES to finite-temperature simulation, which is a more challenging task for quantum lattice models. For optimizing the QES's of 1D systems, we employ the infinite DMRG (iDMRG) algorithm \cite{W92DMRG,W93DMRG}. For the QES's of 2D systems, the $\hat{\mathcal{H}}^{[i,n]}$'s are optimized by using a variant of iDMRG that is developed to simulate the infinite-size models on Bethe lattices \cite{LCP00TreeDMRG,NC13TTN}. In this sense, the environment in the regular lattice (such as honeycomb lattice) is approximated by that of the Bethe lattice. Note that such an approximation can be understood mathematically by the rank-1 decomposition of higher-order tensors \cite{RXLS13NCD}. For 3D systems, the algorithm we use is similar to that for the 2D systems \cite{RPPSL17AOP3D}. Compared with the 2D one, the difference is that the Bethe lattice is an approximation of the 3D lattice, and the coordination number should be different. In general, the computational complexity scales with $\chi$ polynomially as $O(D^{2\mathcal{C}})$ with $\mathcal{C}$ the coordination number of the lattice. For the 1D chain, 2D honeycomb lattice, and 3D cubic lattice for instance, the complexity scales as $O(D^{4})$, $O(D^{6})$, and $O(D^{12})$, respectively. Note that the complexity might be slightly reduced if one chooses a proper contraction order.

To numerically simulate the thermodynamic properties of the QES's, we use the LTRG algorithm. LTRG was originally proposed to simulate the thermodynamic properties of infinite-size 1D systems \cite{LRGZXY+11LTRG}. The idea is to implement the imaginary-time evolution based on the so-called matrix product operator (MPO) form \cite{VGC04MPDO,ZV04MPO,CDV08MPOLR,PMCV10MPO, FND10MPO2D,BKTMWH17FT1D,GIK17FT1D}. In the 1D cases, the QES's are finite chains. Thus, the LTRG is modified on such systems to do imaginary-time evolutions with finite-size MPO's. In the 2D cases, the QES's are finite-size 2D models. We take the idea of 2D DMRG algorithm \cite{WS98tjDMRG,XLS01DMRG2D,SW12DMRG2DRev} to do the simulations. Specifically speaking, a 1D zig-zag path is chosen to cover the finite-size 2D lattice, and the MPO is defined along such a path. The same as the 2D DMRG, the trade-off of this trick is that some nearest-neighbor couplings in the original 2D lattice become long-range in the 1D path. To consider these long-range couplings and capture the (purified) entanglement of the 2D system by the 1D MPO (in fact a thermal state), one will have to face a higher computational complexity and need larger bond dimensions in the LTRG simulations compared with the 1D cases. Note that similar idea was used in Refs. \cite{CDV08MPOLR} and \cite{FND10MPO2D} to construct 2D Hamiltonians instead of finite-temperature thermal states. Our data show that reliable results for small 2D systems (e.g., of O(10) sites) are within the reach of our current computational resources by LTRG.

For the infinite-size XY chain, we use the exact solution by Lieb \textit{et al} \cite{LSM61exact}. For the infinite-size Heisenberg model on honeycomb lattice, we use the QMC algorithm, where the size is extrapolated to infinity. In detail, we use the continuous-time worldline QMC method with ``worm'' update, which was first developed by Prokof’ev and co-workers \cite{PST96QMC, PST98QMC}. The ``worm'' update can efficiently treat the critical slowing-down problem, and there is no Trotter error caused by imaginary-time discretization \cite{KPST99QMC, XHZS+11QMC}.

\section{Experimental implementations}

The approach discussed in this paper is particularly suitable for experimental implementations. For the spin models considered, one needs small-size systems to realize ``simple'' two-body interactions in the ``small'' bulk, and less complicated two-body interactions at the boundary. The state-of-the-art experimental platforms can access about O(10) sites and be used to construct the QES's. According to our numerical simulations, the QES's will largely reduce the finite-size effects that appear in the normal models (without the $\hat{\mathcal{H}}$'s) of such sizes.
\begin{itemize}
\item{\bf Ultracold ions} These systems allow for local control of interactions and thus for designing sophisticated spin models, both in continuous time (cf. \cite{zhang2017observation} with 54 qubits) as well as in digital approach (cf. \cite{friis2018observation} with 20 qubits).
\item{\bf Rydberg atoms} These systems have similar possibilities (cf. \cite{bernien2017probing} with 51 atoms), and can mimic spin systems with long range interactions with local control of interactions.
\item{\bf Ultracold atoms in optical lattices} These systems are better suitable to simulate Hubbard models (cf. \cite{mazurenko2017cold} with about 80 sites). Achieving local control of interactions and tunnelings in these systems is more trick, but can be achieved by local control of magnetic fields and thus Feshbach resonances, or appropriate laser induced hopping and lattice shaking (for some ideas in the context of realising dynamics in curved space via the local control of hoping, see \cite{RTLC17coldatom}).
\item{\bf Superconducting qubits / existing quantum processors} Finally, one could use existing quantum simulators, such as those offered by D-Wave, Google, IMB, Microsoft, etc. \cite{websitesqcompute}, adapting their architectures to the problems of interest, or mapping the problems of interest onto the available architecture.
\end{itemize}

\section{Summary and prospective}

The main contribution of our work is to offer a novel way to model quantum simulators (dubbed as QES) for simulating the thermodynamic properties of infinite-size 1D, 2D and 3D quantum lattice models. Let us try to better understand QES from different points of view. In the algorithmic sense, our work can be used as a finite-temperature approach. An essential difference compared with the existing finite-temperature approaches is that a well-defined finite-size Hamiltonian [Eq. (\ref{eq-hamilt})] is constructed to mimic the targeted infinite-size model. The finite-size Hamiltonian is built in a way similar to the effective Hamiltonians in the numerical renormalization group (NRG) methods (particularly DMRG). We extend the idea of building effective Hamiltonians in the NRG methods by going to higher dimensions and most importantly here from zero to finite temperatures. Moreover, this approach can readily be applied to simulate other quantum lattice models, such as bosons or fermions.

In the theoretical sense, what we are actually doing with QES is to use a finite number of eigenstates defined by $\hat{H}_{QES}$ to reproduce the (reduced) partitioning of an infinite number of eigenstates. In other words, a finite-size canonical ensemble is constructed so that its reduced density matrix of a sub-system accurately mimics that of the infinite-size canonical ensemble. Let us call the system \textit{``entanglement-simutable''} if such a finite-size canonical ensemble can be found. Our work poses a new question: how to theoretically judge whether a given infinite-size system is entanglement-simutable or not. In principle, the QES can accurately simulate the physics in equilibrium as long as the correlation and structure (only for 2D cases) errors are well controlled. For those out of equilibrium, the simulation could be reliable when the time is much shorter than the dynamic correlation length that the QES captures. To simulate the long-time dynamics by a QES, the time-dependent optimization of the physical-bath interactions should work (see the work by Daley \textit{et al} \cite{DKSV04timeHeff} as an inspiring example for 1D cases).

Supplemental to the entanglement-bath picture, the bath Hamiltonian can be considered as a special boundary condition (BC). Normally, a finite-size model can have open or periodic BC. The bath Hamiltonian defines a BC that mimics the infinite-size environment. In 1D quantum systems, such an infinite BC (IBC) was suggested by Phien \textit{et al} \cite{PVM12InfBound} to simulate the time evolutions. In 2D and 3D quantum systems, a similar IBC was suggested to simulate the ground states \cite{RPPSL17AOP3D}.

In the experimental sense, the QES possesses high feasibility of practically realizing it by cold atoms or other platforms. Only O(10) sites are needed to implement accurate simulations. The total Hamiltonian of a QES has a simple form [see Eq. (\ref{eq-hamilt})]. The bulk only consists of the physical interactions of the original model, and the physical-bath interactions on the boundary are only two-body and nearest-neighbor. Besides quantum simulations, the present work also provides a promising way to purposefully realize many-body features in small devices. Suppose that one finds an useful phase by numerically simulating an infinite-size Hamiltonian, but it is impossible at the moment to find a realistic material that is described by such a Hamiltonian. Then one can consider to construct the QES and realize the targeted phase in a O(10)-site device.

In the future, the accuracy of the QES can be further improved. One possible way is to optimize the physical-bath interactions depending on the temperature by, e.g., the imaginary-time sweeping algorithm \cite{RXLS13NCD}. But, there will be no Hamiltonian like Eq. (\ref{eq-hamilt}) since different temperatures give different Hamiltonians. To compute the thermodynamics of a QES more accurately by classical computers, one may employ the series expansion algorithm \cite{CLCL17TNexpand} that is Trotter-error-free. For 2D simulations, one may generalize the pure finite-size projected entangled pair states \cite{LCB14FinitePEPS} to thermal states, where the TN ansatz will respect more the 2D nature of the model. The so-called full update algorithms (e.g., iTEBD \cite{V07iTEBD, JOVVC08PEPS}, CTMRG \cite{NO96CTMRG0}, etc.) can be employed in the same spirit to optimize the Hamiltonians of the QES's in order to better mimic the environment by the bath sites (for instance by introducing the bath-bath interactions).

{\R{The current approach is to handle spins systems, which can be readily generalized to bosonic models. The eventual goal of QES is to deal with the interacting electrons and the realistic materials, where we have the famous dynamic mean-field theory (DMFT) (see \cite{AGPTW10DMF} and two reviews \cite{GKKR96DMFTRev, KSHO+06DMFTrev} for example) and density-matrix embedding theory (DMET) \cite{KC12DMET, KC13DMET} to compete with. These methods share several resemblances with QES. For instance, both reduce the complex system to a much simpler one (few-body model and impurity model for QES and DMFT/DMET, respectively), and both need an eigen-solver (such as DMRG) to solve the reduced models. One advantage of DMFT/DMET is that they can be readily applied to the realistic materials, while the QES and most of the TN methods are restricted to the lattice models (except for some recent progress in the continuous field theories, e.g., \cite{VC10cMPS, JBHOV15cTNS}). However, from the experience of the TN methods, the QES should better consider the strongly-correlated effects (such as quantum entanglement). It is also worth mentioning that the TN methods (such as DMRG and the TN state representations) has been combined with DMFT as efficient eigen-solvers (for example \cite{GHR04DMFTwithDMRG, BZTAE17TNDMFT}). We expect that QES would provide novel paths to further hybridize TN and DMFT/DMET for simulating the interacting electrons with higher efficiency and flexibility.}}

\section*{Acknowledgments}

S.J.R. thanks Tao Xiang for stimulating discussions. S.J.R. acknowledges Fundaci\'o Catalunya - La Pedrera $\cdot$ Ignacio Cirac Program Chair, Beijing Natural Science Foundation (No. 1192005  and Z180013), and Foundation of Beijing Education Committees under Grants No. KZ201810028043. M.L. acknowledges the Spanish Ministry MINECO (National Plan 15 Grant: FISICATEAMO No. FIS2016-79508-P, SEVERO OCHOA No. SEV-2015-0522, FPI), European Social Fund, Fundació Cellex, Generalitat de Catalunya (AGAUR Grant No. 2017 SGR 1341 and CERCA/Program), ERC AdG OSYRIS, EU FETPRO QUIC, and the National Science Centre, Poland-Symfonia Grant No. 2016/20/W/ST4/00314. B.X. is supported by NSFC Grant No. 11774300 and No. 11647316. S.J.R., C.P., and G.S. are supported by NSFC (Grant No. 11834014). C.P., and G.S. are supported by the National Key R\&D Program of China (Grant No. 2018YFA0305800), and the Strategic Priority Research Program of the Chinese Academy of Sciences (Grant No. XDB07010100, XDB28000000).  


\begin{thebibliography}{95}%
\makeatletter
\providecommand \@ifxundefined [1]{%
 \@ifx{#1\undefined}
}%
\providecommand \@ifnum [1]{%
 \ifnum #1\expandafter \@firstoftwo
 \else \expandafter \@secondoftwo
 \fi
}%
\providecommand \@ifx [1]{%
 \ifx #1\expandafter \@firstoftwo
 \else \expandafter \@secondoftwo
 \fi
}%
\providecommand \natexlab [1]{#1}%
\providecommand \enquote  [1]{``#1''}%
\providecommand \bibnamefont  [1]{#1}%
\providecommand \bibfnamefont [1]{#1}%
\providecommand \citenamefont [1]{#1}%
\providecommand \href@noop [0]{\@secondoftwo}%
\providecommand \href [0]{\begingroup \@sanitize@url \@href}%
\providecommand \@href[1]{\@@startlink{#1}\@@href}%
\providecommand \@@href[1]{\endgroup#1\@@endlink}%
\providecommand \@sanitize@url [0]{\catcode `\\12\catcode `\$12\catcode
  `\&12\catcode `\#12\catcode `\^12\catcode `\_12\catcode `\%12\relax}%
\providecommand \@@startlink[1]{}%
\providecommand \@@endlink[0]{}%
\providecommand \url  [0]{\begingroup\@sanitize@url \@url }%
\providecommand \@url [1]{\endgroup\@href {#1}{\urlprefix }}%
\providecommand \urlprefix  [0]{URL }%
\providecommand \Eprint [0]{\href }%
\providecommand \doibase [0]{http://dx.doi.org/}%
\providecommand \selectlanguage [0]{\@gobble}%
\providecommand \bibinfo  [0]{\@secondoftwo}%
\providecommand \bibfield  [0]{\@secondoftwo}%
\providecommand \translation [1]{[#1]}%
\providecommand \BibitemOpen [0]{}%
\providecommand \bibitemStop [0]{}%
\providecommand \bibitemNoStop [0]{.\EOS\space}%
\providecommand \EOS [0]{\spacefactor3000\relax}%
\providecommand \BibitemShut  [1]{\csname bibitem#1\endcsname}%
\let\auto@bib@innerbib\@empty
\bibitem [{\citenamefont {Ravent{\'o}s}\ \emph {et~al.}(2017)\citenamefont
  {Ravent{\'o}s}, \citenamefont {Gra{\ss}}, \citenamefont {Lewenstein},\ and\
  \citenamefont {Juli{\'a}-D{\'\i}az}}]{raventos2017cold}%
  \BibitemOpen
  \bibfield  {author} {\bibinfo {author} {\bibfnamefont {D.}~\bibnamefont
  {Ravent{\'o}s}}, \bibinfo {author} {\bibfnamefont {T.}~\bibnamefont
  {Gra{\ss}}}, \bibinfo {author} {\bibfnamefont {M.}~\bibnamefont
  {Lewenstein}}, \ and\ \bibinfo {author} {\bibfnamefont {B.}~\bibnamefont
  {Juli{\'a}-D{\'\i}az}},\ }\href@noop {} {\bibfield  {journal} {\bibinfo
  {journal} {Journal of Physics B: Atomic, Molecular and Optical Physics}\
  }\textbf {\bibinfo {volume} {50}},\ \bibinfo {pages} {113001} (\bibinfo
  {year} {2017})}\BibitemShut {NoStop}%
\bibitem [{\citenamefont {Ceperley}\ and\ \citenamefont
  {Alder}(1986)}]{CA86QMCrev}%
  \BibitemOpen
  \bibfield  {author} {\bibinfo {author} {\bibfnamefont {D.}~\bibnamefont
  {Ceperley}}\ and\ \bibinfo {author} {\bibfnamefont {B.}~\bibnamefont
  {Alder}},\ }\href {\doibase 10.1126/science.231.4738.555} {\bibfield
  {journal} {\bibinfo  {journal} {Science}\ }\textbf {\bibinfo {volume}
  {231}},\ \bibinfo {pages} {555} (\bibinfo {year} {1986})}\BibitemShut
  {NoStop}%
\bibitem [{\citenamefont {Foulkes}\ \emph {et~al.}(2001)\citenamefont
  {Foulkes}, \citenamefont {Mitas}, \citenamefont {Needs},\ and\ \citenamefont
  {Rajagopal}}]{FMNR01QMCrev}%
  \BibitemOpen
  \bibfield  {author} {\bibinfo {author} {\bibfnamefont {W.~M.~C.}\
  \bibnamefont {Foulkes}}, \bibinfo {author} {\bibfnamefont {L.}~\bibnamefont
  {Mitas}}, \bibinfo {author} {\bibfnamefont {R.~J.}\ \bibnamefont {Needs}}, \
  and\ \bibinfo {author} {\bibfnamefont {G.}~\bibnamefont {Rajagopal}},\ }\href
  {\doibase 10.1103/RevModPhys.73.33} {\bibfield  {journal} {\bibinfo
  {journal} {Rev. Mod. Phys.}\ }\textbf {\bibinfo {volume} {73}},\ \bibinfo
  {pages} {33} (\bibinfo {year} {2001})}\BibitemShut {NoStop}%
\bibitem [{\citenamefont {White}(1992)}]{W92DMRG}%
  \BibitemOpen
  \bibfield  {author} {\bibinfo {author} {\bibfnamefont {S.~R.}\ \bibnamefont
  {White}},\ }\href@noop {} {\bibfield  {journal} {\bibinfo  {journal} {Phys.
  Rev. Lett.}\ }\textbf {\bibinfo {volume} {69}},\ \bibinfo {pages} {2863}
  (\bibinfo {year} {1992})}\BibitemShut {NoStop}%
\bibitem [{\citenamefont {White}(1993)}]{W93DMRG}%
  \BibitemOpen
  \bibfield  {author} {\bibinfo {author} {\bibfnamefont {S.~R.}\ \bibnamefont
  {White}},\ }\href {\doibase 10.1103/PhysRevB.48.10345} {\bibfield  {journal}
  {\bibinfo  {journal} {Phys. Rev. B}\ }\textbf {\bibinfo {volume} {48}},\
  \bibinfo {pages} {10345} (\bibinfo {year} {1993})}\BibitemShut {NoStop}%
\bibitem [{\citenamefont {Verstraete}\ \emph {et~al.}(2008)\citenamefont
  {Verstraete}, \citenamefont {Murg},\ and\ \citenamefont
  {Cirac}}]{VMC08MPSPEPSRev}%
  \BibitemOpen
  \bibfield  {author} {\bibinfo {author} {\bibfnamefont {F.}~\bibnamefont
  {Verstraete}}, \bibinfo {author} {\bibfnamefont {V.}~\bibnamefont {Murg}}, \
  and\ \bibinfo {author} {\bibfnamefont {J.~I.}\ \bibnamefont {Cirac}},\
  }\href@noop {} {\bibfield  {journal} {\bibinfo  {journal} {Advances in
  Physics}\ }\textbf {\bibinfo {volume} {57}},\ \bibinfo {pages} {143}
  (\bibinfo {year} {2008})}\BibitemShut {NoStop}%
\bibitem [{\citenamefont {Cirac}\ and\ \citenamefont
  {Verstraete}(2009)}]{CV09TNSRev}%
  \BibitemOpen
  \bibfield  {author} {\bibinfo {author} {\bibfnamefont {J.~I.}\ \bibnamefont
  {Cirac}}\ and\ \bibinfo {author} {\bibfnamefont {F.}~\bibnamefont
  {Verstraete}},\ }\href@noop {} {\bibfield  {journal} {\bibinfo  {journal} {J.
  Phys. A: Math. Theor.}\ }\textbf {\bibinfo {volume} {42}},\ \bibinfo {pages}
  {504004} (\bibinfo {year} {2009})}\BibitemShut {NoStop}%
\bibitem [{\citenamefont {Schollw\"{o}ck}(2011)}]{S11DMRGRev}%
  \BibitemOpen
  \bibfield  {author} {\bibinfo {author} {\bibfnamefont {U.}~\bibnamefont
  {Schollw\"{o}ck}},\ }\href@noop {} {\bibfield  {journal} {\bibinfo  {journal}
  {Ann. Phys.}\ }\textbf {\bibinfo {volume} {326}},\ \bibinfo {pages} {96}
  (\bibinfo {year} {2011})}\BibitemShut {NoStop}%
\bibitem [{\citenamefont {Or\'{u}s}(2014)}]{O14TNSRev}%
  \BibitemOpen
  \bibfield  {author} {\bibinfo {author} {\bibfnamefont {R.}~\bibnamefont
  {Or\'{u}s}},\ }\href@noop {} {\bibfield  {journal} {\bibinfo  {journal} {Ann.
  Phys.}\ }\textbf {\bibinfo {volume} {349}},\ \bibinfo {pages} {117} (\bibinfo
  {year} {2014})}\BibitemShut {NoStop}%
\bibitem [{\citenamefont {Haegeman}\ and\ \citenamefont
  {Verstraete}(2017)}]{HV17TMTNrev}%
  \BibitemOpen
  \bibfield  {author} {\bibinfo {author} {\bibfnamefont {J.}~\bibnamefont
  {Haegeman}}\ and\ \bibinfo {author} {\bibfnamefont {F.}~\bibnamefont
  {Verstraete}},\ }\href {\doibase 10.1146/annurev-conmatphys-031016-025507}
  {\bibfield  {journal} {\bibinfo  {journal} {Annual Review of Condensed Matter
  Physics}\ }\textbf {\bibinfo {volume} {8}},\ \bibinfo {pages} {355} (\bibinfo
  {year} {2017})}\BibitemShut {NoStop}%
\bibitem [{\citenamefont {Ran}\ \emph {et~al.}(2017{\natexlab{a}})\citenamefont
  {Ran}, \citenamefont {Tirrito}, \citenamefont {Peng}, \citenamefont {Chen},
  \citenamefont {Su},\ and\ \citenamefont {Lewenstein}}]{RTPC+17TNrev}%
  \BibitemOpen
  \bibfield  {author} {\bibinfo {author} {\bibfnamefont {S.-J.}\ \bibnamefont
  {Ran}}, \bibinfo {author} {\bibfnamefont {E.}~\bibnamefont {Tirrito}},
  \bibinfo {author} {\bibfnamefont {C.}~\bibnamefont {Peng}}, \bibinfo {author}
  {\bibfnamefont {X.}~\bibnamefont {Chen}}, \bibinfo {author} {\bibfnamefont
  {G.}~\bibnamefont {Su}}, \ and\ \bibinfo {author} {\bibfnamefont
  {M.}~\bibnamefont {Lewenstein}},\ }\href@noop {} {\bibfield  {journal}
  {\bibinfo  {journal} {arXiv:1708.09213}\ } (\bibinfo {year}
  {2017}{\natexlab{a}})}\BibitemShut {NoStop}%
\bibitem [{\citenamefont {Mila}(2000)}]{M00QSLrev}%
  \BibitemOpen
  \bibfield  {author} {\bibinfo {author} {\bibfnamefont {F.}~\bibnamefont
  {Mila}},\ }\href {http://stacks.iop.org/0143-0807/21/i=6/a=302} {\bibfield
  {journal} {\bibinfo  {journal} {European Journal of Physics}\ }\textbf
  {\bibinfo {volume} {21}},\ \bibinfo {pages} {499} (\bibinfo {year}
  {2000})}\BibitemShut {NoStop}%
\bibitem [{\citenamefont {Balents}(2010)}]{B10QSLRev}%
  \BibitemOpen
  \bibfield  {author} {\bibinfo {author} {\bibfnamefont {L.}~\bibnamefont
  {Balents}},\ }\href@noop {} {\bibfield  {journal} {\bibinfo  {journal}
  {Nature}\ }\textbf {\bibinfo {volume} {464}},\ \bibinfo {pages} {199}
  (\bibinfo {year} {2010})}\BibitemShut {NoStop}%
\bibitem [{\citenamefont {White}(2012)}]{W12QSL}%
  \BibitemOpen
  \bibfield  {author} {\bibinfo {author} {\bibfnamefont {S.~R.}\ \bibnamefont
  {White}},\ }\href@noop {} {\bibfield  {journal} {\bibinfo  {journal} {Nature
  Physics}\ }\textbf {\bibinfo {volume} {8}},\ \bibinfo {pages} {863} (\bibinfo
  {year} {2012})}\BibitemShut {NoStop}%
\bibitem [{\citenamefont {Savary}\ and\ \citenamefont
  {Balents}(2017)}]{SB17QSLRev}%
  \BibitemOpen
  \bibfield  {author} {\bibinfo {author} {\bibfnamefont {L.}~\bibnamefont
  {Savary}}\ and\ \bibinfo {author} {\bibfnamefont {L.}~\bibnamefont
  {Balents}},\ }\href@noop {} {\bibfield  {journal} {\bibinfo  {journal}
  {Reports on Progress in Physics}\ }\textbf {\bibinfo {volume} {80}},\
  \bibinfo {pages} {016502} (\bibinfo {year} {2017})}\BibitemShut {NoStop}%
\bibitem [{\citenamefont {Or\'{u}s}(2012)}]{O12CTMRG}%
  \BibitemOpen
  \bibfield  {author} {\bibinfo {author} {\bibfnamefont {R.}~\bibnamefont
  {Or\'{u}s}},\ }\href@noop {} {\bibfield  {journal} {\bibinfo  {journal}
  {Phys. Rev. B}\ }\textbf {\bibinfo {volume} {85}},\ \bibinfo {pages} {205117}
  (\bibinfo {year} {2012})}\BibitemShut {NoStop}%
\bibitem [{\citenamefont {Ran}\ \emph {et~al.}(2012)\citenamefont {Ran},
  \citenamefont {Li}, \citenamefont {Xi}, \citenamefont {Zhang},\ and\
  \citenamefont {Su}}]{RLXZS12ODTNS}%
  \BibitemOpen
  \bibfield  {author} {\bibinfo {author} {\bibfnamefont {S.-J.}\ \bibnamefont
  {Ran}}, \bibinfo {author} {\bibfnamefont {W.}~\bibnamefont {Li}}, \bibinfo
  {author} {\bibfnamefont {B.}~\bibnamefont {Xi}}, \bibinfo {author}
  {\bibfnamefont {Z.}~\bibnamefont {Zhang}}, \ and\ \bibinfo {author}
  {\bibfnamefont {G.}~\bibnamefont {Su}},\ }\href@noop {} {\bibfield  {journal}
  {\bibinfo  {journal} {Phys. Rev. B}\ }\textbf {\bibinfo {volume} {86}},\
  \bibinfo {pages} {134429} (\bibinfo {year} {2012})}\BibitemShut {NoStop}%
\bibitem [{\citenamefont {Czarnik}\ \emph {et~al.}(2012)\citenamefont
  {Czarnik}, \citenamefont {Cincio},\ and\ \citenamefont
  {Dziarmaga}}]{CCD12FTPEPS}%
  \BibitemOpen
  \bibfield  {author} {\bibinfo {author} {\bibfnamefont {P.}~\bibnamefont
  {Czarnik}}, \bibinfo {author} {\bibfnamefont {L.}~\bibnamefont {Cincio}}, \
  and\ \bibinfo {author} {\bibfnamefont {J.}~\bibnamefont {Dziarmaga}},\
  }\href@noop {} {\bibfield  {journal} {\bibinfo  {journal} {Phys. Rev. B}\
  }\textbf {\bibinfo {volume} {86}},\ \bibinfo {pages} {245101} (\bibinfo
  {year} {2012})}\BibitemShut {NoStop}%
\bibitem [{\citenamefont {Ran}\ \emph {et~al.}(2013)\citenamefont {Ran},
  \citenamefont {Xi}, \citenamefont {Liu},\ and\ \citenamefont
  {Su}}]{RXLS13NCD}%
  \BibitemOpen
  \bibfield  {author} {\bibinfo {author} {\bibfnamefont {S.-J.}\ \bibnamefont
  {Ran}}, \bibinfo {author} {\bibfnamefont {B.}~\bibnamefont {Xi}}, \bibinfo
  {author} {\bibfnamefont {T.-Y.}\ \bibnamefont {Liu}}, \ and\ \bibinfo
  {author} {\bibfnamefont {G.}~\bibnamefont {Su}},\ }\href@noop {} {\bibfield
  {journal} {\bibinfo  {journal} {Phys. Rev. B}\ }\textbf {\bibinfo {volume}
  {88}},\ \bibinfo {pages} {064407} (\bibinfo {year} {2013})}\BibitemShut
  {NoStop}%
\bibitem [{\citenamefont {Czarnik}\ and\ \citenamefont
  {Dziarmaga}(2015{\natexlab{a}})}]{CD15TPO}%
  \BibitemOpen
  \bibfield  {author} {\bibinfo {author} {\bibfnamefont {P.}~\bibnamefont
  {Czarnik}}\ and\ \bibinfo {author} {\bibfnamefont {J.}~\bibnamefont
  {Dziarmaga}},\ }\href {\doibase 10.1103/PhysRevB.92.035152} {\bibfield
  {journal} {\bibinfo  {journal} {Phys. Rev. B}\ }\textbf {\bibinfo {volume}
  {92}},\ \bibinfo {pages} {035152} (\bibinfo {year}
  {2015}{\natexlab{a}})}\BibitemShut {NoStop}%
\bibitem [{\citenamefont {Czarnik}\ and\ \citenamefont
  {Dziarmaga}(2015{\natexlab{b}})}]{CD15FTPEPS}%
  \BibitemOpen
  \bibfield  {author} {\bibinfo {author} {\bibfnamefont {P.}~\bibnamefont
  {Czarnik}}\ and\ \bibinfo {author} {\bibfnamefont {J.}~\bibnamefont
  {Dziarmaga}},\ }\href {\doibase 10.1103/PhysRevB.92.035120} {\bibfield
  {journal} {\bibinfo  {journal} {Phys. Rev. B}\ }\textbf {\bibinfo {volume}
  {92}},\ \bibinfo {pages} {035120} (\bibinfo {year}
  {2015}{\natexlab{b}})}\BibitemShut {NoStop}%
\bibitem [{\citenamefont {Czarnik}\ \emph
  {et~al.}(2016{\natexlab{a}})\citenamefont {Czarnik}, \citenamefont
  {Dziarmaga},\ and\ \citenamefont {Ole\'{s}}}]{CDO16TPO}%
  \BibitemOpen
  \bibfield  {author} {\bibinfo {author} {\bibfnamefont {P.}~\bibnamefont
  {Czarnik}}, \bibinfo {author} {\bibfnamefont {J.}~\bibnamefont {Dziarmaga}},
  \ and\ \bibinfo {author} {\bibfnamefont {A.~M.}\ \bibnamefont {Ole\'{s}}},\
  }\href {\doibase 10.1103/PhysRevB.93.184410} {\bibfield  {journal} {\bibinfo
  {journal} {Phys. Rev. B}\ }\textbf {\bibinfo {volume} {93}},\ \bibinfo
  {pages} {184410} (\bibinfo {year} {2016}{\natexlab{a}})}\BibitemShut
  {NoStop}%
\bibitem [{\citenamefont {Czarnik}\ \emph
  {et~al.}(2016{\natexlab{b}})\citenamefont {Czarnik}, \citenamefont {Rams},\
  and\ \citenamefont {Dziarmaga}}]{CRD16TPO}%
  \BibitemOpen
  \bibfield  {author} {\bibinfo {author} {\bibfnamefont {P.}~\bibnamefont
  {Czarnik}}, \bibinfo {author} {\bibfnamefont {M.~M.}\ \bibnamefont {Rams}}, \
  and\ \bibinfo {author} {\bibfnamefont {J.}~\bibnamefont {Dziarmaga}},\ }\href
  {\doibase 10.1103/PhysRevB.94.235142} {\bibfield  {journal} {\bibinfo
  {journal} {Phys. Rev. B}\ }\textbf {\bibinfo {volume} {94}},\ \bibinfo
  {pages} {235142} (\bibinfo {year} {2016}{\natexlab{b}})}\BibitemShut
  {NoStop}%
\bibitem [{\citenamefont {Czarnik}\ \emph {et~al.}(2017)\citenamefont
  {Czarnik}, \citenamefont {Dziarmaga},\ and\ \citenamefont
  {Olei\'{s}}}]{CDO17TPOQMC}%
  \BibitemOpen
  \bibfield  {author} {\bibinfo {author} {\bibfnamefont {P.}~\bibnamefont
  {Czarnik}}, \bibinfo {author} {\bibfnamefont {J.}~\bibnamefont {Dziarmaga}},
  \ and\ \bibinfo {author} {\bibfnamefont {A.~M.}\ \bibnamefont {Olei\'{s}}},\
  }\href {\doibase 10.1103/PhysRevB.96.014420} {\bibfield  {journal} {\bibinfo
  {journal} {Phys. Rev. B}\ }\textbf {\bibinfo {volume} {96}},\ \bibinfo
  {pages} {014420} (\bibinfo {year} {2017})}\BibitemShut {NoStop}%
\bibitem [{\citenamefont {Banerjee}\ \emph {et~al.}(2016)\citenamefont
  {Banerjee}, \citenamefont {Bridges}, \citenamefont {Yan}, \citenamefont
  {Aczel}, \citenamefont {Li}, \citenamefont {Stone}, \citenamefont {Granroth},
  \citenamefont {Lumsden}, \citenamefont {Yiu}, \citenamefont {Knolle},
  \citenamefont {Bhattacharjee}, \citenamefont {Kovrizhin}, \citenamefont
  {Moessner}, \citenamefont {Tennant}, \citenamefont {Mandrus},\ and\
  \citenamefont {Nagler}}]{BBYA+15QSLexp}%
  \BibitemOpen
  \bibfield  {author} {\bibinfo {author} {\bibfnamefont {A.}~\bibnamefont
  {Banerjee}}, \bibinfo {author} {\bibfnamefont {C.~A.}\ \bibnamefont
  {Bridges}}, \bibinfo {author} {\bibfnamefont {J.~Q.}\ \bibnamefont {Yan}},
  \bibinfo {author} {\bibfnamefont {A.~A.}\ \bibnamefont {Aczel}}, \bibinfo
  {author} {\bibfnamefont {L.}~\bibnamefont {Li}}, \bibinfo {author}
  {\bibfnamefont {M.~B.}\ \bibnamefont {Stone}}, \bibinfo {author}
  {\bibfnamefont {G.~E.}\ \bibnamefont {Granroth}}, \bibinfo {author}
  {\bibfnamefont {M.~D.}\ \bibnamefont {Lumsden}}, \bibinfo {author}
  {\bibfnamefont {Y.}~\bibnamefont {Yiu}}, \bibinfo {author} {\bibfnamefont
  {J.}~\bibnamefont {Knolle}}, \bibinfo {author} {\bibfnamefont
  {S.}~\bibnamefont {Bhattacharjee}}, \bibinfo {author} {\bibfnamefont {D.~L.}\
  \bibnamefont {Kovrizhin}}, \bibinfo {author} {\bibfnamefont {R.}~\bibnamefont
  {Moessner}}, \bibinfo {author} {\bibfnamefont {D.~A.}\ \bibnamefont
  {Tennant}}, \bibinfo {author} {\bibfnamefont {D.~G.}\ \bibnamefont
  {Mandrus}}, \ and\ \bibinfo {author} {\bibfnamefont {S.~E.}\ \bibnamefont
  {Nagler}},\ }\href {\doibase 10.1038/nmat4604
  https://www.nature.com/articles/nmat4604#supplementary-information}
  {\bibfield  {journal} {\bibinfo  {journal} {Nature Materials}\ }\textbf
  {\bibinfo {volume} {15}},\ \bibinfo {pages} {733} (\bibinfo {year}
  {2016})}\BibitemShut {NoStop}%
\bibitem [{\citenamefont {Shen}\ \emph {et~al.}(2016)\citenamefont {Shen},
  \citenamefont {Li}, \citenamefont {Wo}, \citenamefont {Li}, \citenamefont
  {Shen}, \citenamefont {Pan}, \citenamefont {Wang}, \citenamefont {Walker},
  \citenamefont {Steffens}, \citenamefont {Boehm}, \citenamefont {Hao},
  \citenamefont {Quintero-Castro}, \citenamefont {Harriger}, \citenamefont
  {Frontzek}, \citenamefont {Hao}, \citenamefont {Meng}, \citenamefont {Zhang},
  \citenamefont {Chen},\ and\ \citenamefont {Zhao}}]{LSWL+16QSLexp}%
  \BibitemOpen
  \bibfield  {author} {\bibinfo {author} {\bibfnamefont {Y.}~\bibnamefont
  {Shen}}, \bibinfo {author} {\bibfnamefont {Y.-D.}\ \bibnamefont {Li}},
  \bibinfo {author} {\bibfnamefont {H.}~\bibnamefont {Wo}}, \bibinfo {author}
  {\bibfnamefont {Y.}~\bibnamefont {Li}}, \bibinfo {author} {\bibfnamefont
  {S.}~\bibnamefont {Shen}}, \bibinfo {author} {\bibfnamefont {B.}~\bibnamefont
  {Pan}}, \bibinfo {author} {\bibfnamefont {Q.}~\bibnamefont {Wang}}, \bibinfo
  {author} {\bibfnamefont {H.~C.}\ \bibnamefont {Walker}}, \bibinfo {author}
  {\bibfnamefont {P.}~\bibnamefont {Steffens}}, \bibinfo {author}
  {\bibfnamefont {M.}~\bibnamefont {Boehm}}, \bibinfo {author} {\bibfnamefont
  {Y.}~\bibnamefont {Hao}}, \bibinfo {author} {\bibfnamefont {D.~L.}\
  \bibnamefont {Quintero-Castro}}, \bibinfo {author} {\bibfnamefont {L.~W.}\
  \bibnamefont {Harriger}}, \bibinfo {author} {\bibfnamefont {M.~D.}\
  \bibnamefont {Frontzek}}, \bibinfo {author} {\bibfnamefont {L.}~\bibnamefont
  {Hao}}, \bibinfo {author} {\bibfnamefont {S.}~\bibnamefont {Meng}}, \bibinfo
  {author} {\bibfnamefont {Q.}~\bibnamefont {Zhang}}, \bibinfo {author}
  {\bibfnamefont {G.}~\bibnamefont {Chen}}, \ and\ \bibinfo {author}
  {\bibfnamefont {J.}~\bibnamefont {Zhao}},\ }\href {\doibase
  10.1038/nature20614} {\bibfield  {journal} {\bibinfo  {journal} {Nature}\
  }\textbf {\bibinfo {volume} {540}},\ \bibinfo {pages} {559} (\bibinfo {year}
  {2016})}\BibitemShut {NoStop}%
\bibitem [{\citenamefont {Han}\ \emph {et~al.}(2012)\citenamefont {Han},
  \citenamefont {Helton}, \citenamefont {Chu}, \citenamefont {Nocera},
  \citenamefont {Rodriguez-Rivera}, \citenamefont {Broholm},\ and\
  \citenamefont {Lee}}]{HHCNR+12fractionalized}%
  \BibitemOpen
  \bibfield  {author} {\bibinfo {author} {\bibfnamefont {T.-H.}\ \bibnamefont
  {Han}}, \bibinfo {author} {\bibfnamefont {J.~S.}\ \bibnamefont {Helton}},
  \bibinfo {author} {\bibfnamefont {S.~Y.}\ \bibnamefont {Chu}}, \bibinfo
  {author} {\bibfnamefont {D.~G.}\ \bibnamefont {Nocera}}, \bibinfo {author}
  {\bibfnamefont {J.~A.}\ \bibnamefont {Rodriguez-Rivera}}, \bibinfo {author}
  {\bibfnamefont {C.}~\bibnamefont {Broholm}}, \ and\ \bibinfo {author}
  {\bibfnamefont {Y.~S.}\ \bibnamefont {Lee}},\ }\href@noop {} {\bibfield
  {journal} {\bibinfo  {journal} {Nature}\ }\textbf {\bibinfo {volume} {492}},\
  \bibinfo {pages} {406} (\bibinfo {year} {2012})}\BibitemShut {NoStop}%
\bibitem [{\citenamefont {Lewenstein}\ \emph {et~al.}(2012)\citenamefont
  {Lewenstein}, \citenamefont {Sanpera},\ and\ \citenamefont
  {Ahufinger}}]{LSA12SimuBook}%
  \BibitemOpen
  \bibfield  {author} {\bibinfo {author} {\bibfnamefont {M.}~\bibnamefont
  {Lewenstein}}, \bibinfo {author} {\bibfnamefont {A.}~\bibnamefont {Sanpera}},
  \ and\ \bibinfo {author} {\bibfnamefont {V.}~\bibnamefont {Ahufinger}},\
  }\href@noop {} {\emph {\bibinfo {title} {Ultracold atoms in Optical Lattices:
  simulating quantum many body physics}}}\ (\bibinfo  {publisher} {Oxford
  University Press, Oxford},\ \bibinfo {year} {2012})\BibitemShut {NoStop}%
\bibitem [{\citenamefont {Cirac}\ and\ \citenamefont
  {Zoller}(2012)}]{CZ12QsimuRev}%
  \BibitemOpen
  \bibfield  {author} {\bibinfo {author} {\bibfnamefont {J.~I.}\ \bibnamefont
  {Cirac}}\ and\ \bibinfo {author} {\bibfnamefont {P.}~\bibnamefont {Zoller}},\
  }\href@noop {} {\bibfield  {journal} {\bibinfo  {journal} {Nature Physics}\
  }\textbf {\bibinfo {volume} {8}},\ \bibinfo {pages} {264} (\bibinfo {year}
  {2012})}\BibitemShut {NoStop}%
\bibitem [{\citenamefont {Bloch}\ \emph {et~al.}(2012)\citenamefont {Bloch},
  \citenamefont {Dalibard},\ and\ \citenamefont {Nascimbene}}]{BDN12Qsimu}%
  \BibitemOpen
  \bibfield  {author} {\bibinfo {author} {\bibfnamefont {I.}~\bibnamefont
  {Bloch}}, \bibinfo {author} {\bibfnamefont {J.}~\bibnamefont {Dalibard}}, \
  and\ \bibinfo {author} {\bibfnamefont {S.}~\bibnamefont {Nascimbene}},\
  }\href@noop {} {\bibfield  {journal} {\bibinfo  {journal} {Nature Physics}\
  }\textbf {\bibinfo {volume} {8}},\ \bibinfo {pages} {267} (\bibinfo {year}
  {2012})}\BibitemShut {NoStop}%
\bibitem [{\citenamefont {Aspuru-Guzik}\ and\ \citenamefont
  {Walther}(2012)}]{AW12simu}%
  \BibitemOpen
  \bibfield  {author} {\bibinfo {author} {\bibfnamefont {A.}~\bibnamefont
  {Aspuru-Guzik}}\ and\ \bibinfo {author} {\bibfnamefont {P.}~\bibnamefont
  {Walther}},\ }\href {\doibase 10.1038/nphys2253} {\bibfield  {journal}
  {\bibinfo  {journal} {Nature Physics}\ }\textbf {\bibinfo {volume} {8}},\
  \bibinfo {pages} {285} (\bibinfo {year} {2012})}\BibitemShut {NoStop}%
\bibitem [{\citenamefont {Georgescu}\ \emph {et~al.}(2014)\citenamefont
  {Georgescu}, \citenamefont {Ashhab},\ and\ \citenamefont {Nori}}]{GAN14simu}%
  \BibitemOpen
  \bibfield  {author} {\bibinfo {author} {\bibfnamefont {I.~M.}\ \bibnamefont
  {Georgescu}}, \bibinfo {author} {\bibfnamefont {S.}~\bibnamefont {Ashhab}}, \
  and\ \bibinfo {author} {\bibfnamefont {F.}~\bibnamefont {Nori}},\ }\href
  {\doibase 10.1103/RevModPhys.86.153} {\bibfield  {journal} {\bibinfo
  {journal} {Rev. Mod. Phys.}\ }\textbf {\bibinfo {volume} {86}},\ \bibinfo
  {pages} {153} (\bibinfo {year} {2014})}\BibitemShut {NoStop}%
\bibitem [{\citenamefont {Acin}\ \emph {et~al.}(2018)\citenamefont {Acin},
  \citenamefont {Bloch}, \citenamefont {Buhrman}, \citenamefont {Calarco},
  \citenamefont {Eichler}, \citenamefont {Eisert}, \citenamefont {Esteve},
  \citenamefont {Gisin}, \citenamefont {Glaser}, \citenamefont {Jelezko} \emph
  {et~al.}}]{ABBCE+18QTroadmap}%
  \BibitemOpen
  \bibfield  {author} {\bibinfo {author} {\bibfnamefont {A.}~\bibnamefont
  {Acin}}, \bibinfo {author} {\bibfnamefont {I.}~\bibnamefont {Bloch}},
  \bibinfo {author} {\bibfnamefont {H.}~\bibnamefont {Buhrman}}, \bibinfo
  {author} {\bibfnamefont {T.}~\bibnamefont {Calarco}}, \bibinfo {author}
  {\bibfnamefont {C.}~\bibnamefont {Eichler}}, \bibinfo {author} {\bibfnamefont
  {J.}~\bibnamefont {Eisert}}, \bibinfo {author} {\bibfnamefont
  {D.}~\bibnamefont {Esteve}}, \bibinfo {author} {\bibfnamefont
  {N.}~\bibnamefont {Gisin}}, \bibinfo {author} {\bibfnamefont {S.~J.}\
  \bibnamefont {Glaser}}, \bibinfo {author} {\bibfnamefont {F.}~\bibnamefont
  {Jelezko}},  \emph {et~al.},\ }\href@noop {} {\bibfield  {journal} {\bibinfo
  {journal} {New Journal of Physics}\ }\textbf {\bibinfo {volume} {20}},\
  \bibinfo {pages} {080201} (\bibinfo {year} {2018})}\BibitemShut {NoStop}%
\bibitem [{\citenamefont {Davis}\ \emph {et~al.}(1995)\citenamefont {Davis},
  \citenamefont {Mewes}, \citenamefont {Andrews}, \citenamefont {van Druten},
  \citenamefont {Durfee}, \citenamefont {Kurn},\ and\ \citenamefont
  {Ketterle}}]{DMAD95BEC}%
  \BibitemOpen
  \bibfield  {author} {\bibinfo {author} {\bibfnamefont {K.~B.}\ \bibnamefont
  {Davis}}, \bibinfo {author} {\bibfnamefont {M.~O.}\ \bibnamefont {Mewes}},
  \bibinfo {author} {\bibfnamefont {M.~R.}\ \bibnamefont {Andrews}}, \bibinfo
  {author} {\bibfnamefont {N.~J.}\ \bibnamefont {van Druten}}, \bibinfo
  {author} {\bibfnamefont {D.~S.}\ \bibnamefont {Durfee}}, \bibinfo {author}
  {\bibfnamefont {D.~M.}\ \bibnamefont {Kurn}}, \ and\ \bibinfo {author}
  {\bibfnamefont {W.}~\bibnamefont {Ketterle}},\ }\href {\doibase
  10.1103/PhysRevLett.75.3969} {\bibfield  {journal} {\bibinfo  {journal}
  {Phys. Rev. Lett.}\ }\textbf {\bibinfo {volume} {75}},\ \bibinfo {pages}
  {3969} (\bibinfo {year} {1995})}\BibitemShut {NoStop}%
\bibitem [{\citenamefont {Anderson}\ \emph {et~al.}(1995)\citenamefont
  {Anderson}, \citenamefont {Ensher}, \citenamefont {Matthews}, \citenamefont
  {Wieman},\ and\ \citenamefont {Cornell}}]{anderson1995observation}%
  \BibitemOpen
  \bibfield  {author} {\bibinfo {author} {\bibfnamefont {M.~H.}\ \bibnamefont
  {Anderson}}, \bibinfo {author} {\bibfnamefont {J.~R.}\ \bibnamefont
  {Ensher}}, \bibinfo {author} {\bibfnamefont {M.~R.}\ \bibnamefont
  {Matthews}}, \bibinfo {author} {\bibfnamefont {C.~E.}\ \bibnamefont
  {Wieman}}, \ and\ \bibinfo {author} {\bibfnamefont {E.~A.}\ \bibnamefont
  {Cornell}},\ }\href@noop {} {\bibfield  {journal} {\bibinfo  {journal}
  {science}\ }\textbf {\bibinfo {volume} {269}},\ \bibinfo {pages} {198}
  (\bibinfo {year} {1995})}\BibitemShut {NoStop}%
\bibitem [{\citenamefont {Griesmaier}\ \emph {et~al.}(2005)\citenamefont
  {Griesmaier}, \citenamefont {Werner}, \citenamefont {Hensler}, \citenamefont
  {Stuhler},\ and\ \citenamefont {Pfau}}]{GWHSP05BEC}%
  \BibitemOpen
  \bibfield  {author} {\bibinfo {author} {\bibfnamefont {A.}~\bibnamefont
  {Griesmaier}}, \bibinfo {author} {\bibfnamefont {J.}~\bibnamefont {Werner}},
  \bibinfo {author} {\bibfnamefont {S.}~\bibnamefont {Hensler}}, \bibinfo
  {author} {\bibfnamefont {J.}~\bibnamefont {Stuhler}}, \ and\ \bibinfo
  {author} {\bibfnamefont {T.}~\bibnamefont {Pfau}},\ }\href {\doibase
  10.1103/PhysRevLett.94.160401} {\bibfield  {journal} {\bibinfo  {journal}
  {Phys. Rev. Lett.}\ }\textbf {\bibinfo {volume} {94}},\ \bibinfo {pages}
  {160401} (\bibinfo {year} {2005})}\BibitemShut {NoStop}%
\bibitem [{\citenamefont {Friedenauer}\ \emph {et~al.}(2008)\citenamefont
  {Friedenauer}, \citenamefont {Schmitz}, \citenamefont {Glueckert},
  \citenamefont {Porras},\ and\ \citenamefont {Schaetz}}]{FSGPS08qsimuexp}%
  \BibitemOpen
  \bibfield  {author} {\bibinfo {author} {\bibfnamefont {A.}~\bibnamefont
  {Friedenauer}}, \bibinfo {author} {\bibfnamefont {H.}~\bibnamefont
  {Schmitz}}, \bibinfo {author} {\bibfnamefont {J.~T.}\ \bibnamefont
  {Glueckert}}, \bibinfo {author} {\bibfnamefont {D.}~\bibnamefont {Porras}}, \
  and\ \bibinfo {author} {\bibfnamefont {T.}~\bibnamefont {Schaetz}},\ }\href
  {\doibase 10.1038/nphys1032} {\bibfield  {journal} {\bibinfo  {journal}
  {Nature Physics}\ }\textbf {\bibinfo {volume} {4}},\ \bibinfo {pages} {757}
  (\bibinfo {year} {2008})}\BibitemShut {NoStop}%
\bibitem [{\citenamefont {Simon}\ \emph {et~al.}(2011)\citenamefont {Simon},
  \citenamefont {Bakr}, \citenamefont {Ma}, \citenamefont {Tai}, \citenamefont
  {Preiss},\ and\ \citenamefont {Greiner}}]{SBMT+11AFsimu}%
  \BibitemOpen
  \bibfield  {author} {\bibinfo {author} {\bibfnamefont {J.}~\bibnamefont
  {Simon}}, \bibinfo {author} {\bibfnamefont {W.~S.}\ \bibnamefont {Bakr}},
  \bibinfo {author} {\bibfnamefont {R.}~\bibnamefont {Ma}}, \bibinfo {author}
  {\bibfnamefont {M.~E.}\ \bibnamefont {Tai}}, \bibinfo {author} {\bibfnamefont
  {P.~M.}\ \bibnamefont {Preiss}}, \ and\ \bibinfo {author} {\bibfnamefont
  {M.}~\bibnamefont {Greiner}},\ }\href@noop {} {\bibfield  {journal} {\bibinfo
   {journal} {Nature}\ }\textbf {\bibinfo {volume} {472}},\ \bibinfo {pages}
  {307} (\bibinfo {year} {2011})}\BibitemShut {NoStop}%
\bibitem [{\citenamefont {J\"ordens}\ \emph {et~al.}(2008)\citenamefont
  {J\"ordens}, \citenamefont {Strohmaier}, \citenamefont {G\"unter},
  \citenamefont {Moritz},\ and\ \citenamefont {Esslinger}}]{JSGME08qsimu}%
  \BibitemOpen
  \bibfield  {author} {\bibinfo {author} {\bibfnamefont {R.}~\bibnamefont
  {J\"ordens}}, \bibinfo {author} {\bibfnamefont {N.}~\bibnamefont
  {Strohmaier}}, \bibinfo {author} {\bibfnamefont {K.}~\bibnamefont
  {G\"unter}}, \bibinfo {author} {\bibfnamefont {H.}~\bibnamefont {Moritz}}, \
  and\ \bibinfo {author} {\bibfnamefont {T.}~\bibnamefont {Esslinger}},\ }\href
  {\doibase 10.1038/nature07244} {\bibfield  {journal} {\bibinfo  {journal}
  {Nature}\ }\textbf {\bibinfo {volume} {455}},\ \bibinfo {pages} {204}
  (\bibinfo {year} {2008})}\BibitemShut {NoStop}%
\bibitem [{\citenamefont {Chiu}\ \emph {et~al.}(2018)\citenamefont {Chiu},
  \citenamefont {Ji}, \citenamefont {Mazurenko}, \citenamefont {Greif},\ and\
  \citenamefont {Greiner}}]{CJMGG18fermionSimu}%
  \BibitemOpen
  \bibfield  {author} {\bibinfo {author} {\bibfnamefont {C.~S.}\ \bibnamefont
  {Chiu}}, \bibinfo {author} {\bibfnamefont {G.}~\bibnamefont {Ji}}, \bibinfo
  {author} {\bibfnamefont {A.}~\bibnamefont {Mazurenko}}, \bibinfo {author}
  {\bibfnamefont {D.}~\bibnamefont {Greif}}, \ and\ \bibinfo {author}
  {\bibfnamefont {M.}~\bibnamefont {Greiner}},\ }\href {\doibase
  10.1103/PhysRevLett.120.243201} {\bibfield  {journal} {\bibinfo  {journal}
  {Phys. Rev. Lett.}\ }\textbf {\bibinfo {volume} {120}},\ \bibinfo {pages}
  {243201} (\bibinfo {year} {2018})}\BibitemShut {NoStop}%
\bibitem [{\citenamefont {Ran}\ \emph {et~al.}(2017{\natexlab{b}})\citenamefont
  {Ran}, \citenamefont {Piga}, \citenamefont {Peng}, \citenamefont {Su},\ and\
  \citenamefont {Lewenstein}}]{RPPSL17AOP3D}%
  \BibitemOpen
  \bibfield  {author} {\bibinfo {author} {\bibfnamefont {S.-J.}\ \bibnamefont
  {Ran}}, \bibinfo {author} {\bibfnamefont {A.}~\bibnamefont {Piga}}, \bibinfo
  {author} {\bibfnamefont {C.}~\bibnamefont {Peng}}, \bibinfo {author}
  {\bibfnamefont {G.}~\bibnamefont {Su}}, \ and\ \bibinfo {author}
  {\bibfnamefont {M.}~\bibnamefont {Lewenstein}},\ }\href {\doibase
  10.1103/PhysRevB.96.155120} {\bibfield  {journal} {\bibinfo  {journal} {Phys.
  Rev. B}\ }\textbf {\bibinfo {volume} {96}},\ \bibinfo {pages} {155120}
  (\bibinfo {year} {2017}{\natexlab{b}})}\BibitemShut {NoStop}%
\bibitem [{\citenamefont {Lepetit}\ \emph {et~al.}(2000)\citenamefont
  {Lepetit}, \citenamefont {Cousy},\ and\ \citenamefont
  {Pastor}}]{LCP00TreeDMRG}%
  \BibitemOpen
  \bibfield  {author} {\bibinfo {author} {\bibfnamefont {M.-B.}\ \bibnamefont
  {Lepetit}}, \bibinfo {author} {\bibfnamefont {M.}~\bibnamefont {Cousy}}, \
  and\ \bibinfo {author} {\bibfnamefont {G.~M.}\ \bibnamefont {Pastor}},\
  }\href@noop {} {\bibfield  {journal} {\bibinfo  {journal} {Eur. Phys. J. B}\
  }\textbf {\bibinfo {volume} {13}},\ \bibinfo {pages} {421} (\bibinfo {year}
  {2000})}\BibitemShut {NoStop}%
\bibitem [{\citenamefont {Nakatani}\ and\ \citenamefont
  {Chan}(2013)}]{NC13TTN}%
  \BibitemOpen
  \bibfield  {author} {\bibinfo {author} {\bibfnamefont {N.}~\bibnamefont
  {Nakatani}}\ and\ \bibinfo {author} {\bibfnamefont {G.~K.~L.}\ \bibnamefont
  {Chan}},\ }\href@noop {} {\bibfield  {journal} {\bibinfo  {journal} {J. Chem.
  Phys.}\ }\textbf {\bibinfo {volume} {138}},\ \bibinfo {pages} {134113}
  (\bibinfo {year} {2013})}\BibitemShut {NoStop}%
\bibitem [{Note1()}]{Note1}%
  \BibitemOpen
  \bibinfo {note} {The codes for obtaining the physical-bath Hamiltonians
  $\protect \mathaccentV {hat}05E{\protect \mathcal {H}}$ with the necessary
  instructions can be found at \protect \url
  {https://github.com/ranshiju/FT-QES}.}\BibitemShut {Stop}%
\bibitem [{\citenamefont {Li}\ \emph {et~al.}(2011)\citenamefont {Li},
  \citenamefont {Ran}, \citenamefont {Gong}, \citenamefont {Zhao},
  \citenamefont {Xi}, \citenamefont {Ye},\ and\ \citenamefont
  {Su}}]{LRGZXY+11LTRG}%
  \BibitemOpen
  \bibfield  {author} {\bibinfo {author} {\bibfnamefont {W.}~\bibnamefont
  {Li}}, \bibinfo {author} {\bibfnamefont {S.-J.}\ \bibnamefont {Ran}},
  \bibinfo {author} {\bibfnamefont {S.-S.}\ \bibnamefont {Gong}}, \bibinfo
  {author} {\bibfnamefont {Y.}~\bibnamefont {Zhao}}, \bibinfo {author}
  {\bibfnamefont {B.}~\bibnamefont {Xi}}, \bibinfo {author} {\bibfnamefont
  {F.}~\bibnamefont {Ye}}, \ and\ \bibinfo {author} {\bibfnamefont
  {G.}~\bibnamefont {Su}},\ }\href@noop {} {\bibfield  {journal} {\bibinfo
  {journal} {Phys. Rev. Lett.}\ }\textbf {\bibinfo {volume} {106}},\ \bibinfo
  {pages} {127202} (\bibinfo {year} {2011})}\BibitemShut {NoStop}%
\bibitem [{\citenamefont {Lieb}\ \emph {et~al.}(1961)\citenamefont {Lieb},
  \citenamefont {Schultz},\ and\ \citenamefont {Mattis}}]{LSM61exact}%
  \BibitemOpen
  \bibfield  {author} {\bibinfo {author} {\bibfnamefont {E.}~\bibnamefont
  {Lieb}}, \bibinfo {author} {\bibfnamefont {T.}~\bibnamefont {Schultz}}, \
  and\ \bibinfo {author} {\bibfnamefont {D.}~\bibnamefont {Mattis}},\ }\href
  {\doibase https://doi.org/10.1016/0003-4916(61)90115-4} {\bibfield  {journal}
  {\bibinfo  {journal} {Ann. Phys.}\ }\textbf {\bibinfo {volume} {16}},\
  \bibinfo {pages} {407 } (\bibinfo {year} {1961})}\BibitemShut {NoStop}%
\bibitem [{\citenamefont {Tagliacozzo}\ \emph {et~al.}(2008)\citenamefont
  {Tagliacozzo}, \citenamefont {de~Oliveira}, \citenamefont {Iblisdir},\ and\
  \citenamefont {Latorre}}]{TOIL08EntScaling}%
  \BibitemOpen
  \bibfield  {author} {\bibinfo {author} {\bibfnamefont {L.}~\bibnamefont
  {Tagliacozzo}}, \bibinfo {author} {\bibfnamefont {T.}~\bibnamefont
  {de~Oliveira}}, \bibinfo {author} {\bibfnamefont {S.}~\bibnamefont
  {Iblisdir}}, \ and\ \bibinfo {author} {\bibfnamefont {J.~I.}\ \bibnamefont
  {Latorre}},\ }\href {\doibase 10.1103/PhysRevB.78.024410} {\bibfield
  {journal} {\bibinfo  {journal} {Phys. Rev. B}\ }\textbf {\bibinfo {volume}
  {78}},\ \bibinfo {pages} {024410} (\bibinfo {year} {2008})}\BibitemShut
  {NoStop}%
\bibitem [{\citenamefont {Tirrito}\ \emph {et~al.}(2018)\citenamefont
  {Tirrito}, \citenamefont {Tagliacozzo}, \citenamefont {Lewenstein},\ and\
  \citenamefont {Ran}}]{TTLR18tMPS}%
  \BibitemOpen
  \bibfield  {author} {\bibinfo {author} {\bibfnamefont {E.}~\bibnamefont
  {Tirrito}}, \bibinfo {author} {\bibfnamefont {L.}~\bibnamefont
  {Tagliacozzo}}, \bibinfo {author} {\bibfnamefont {M.}~\bibnamefont
  {Lewenstein}}, \ and\ \bibinfo {author} {\bibfnamefont {S.-J.}\ \bibnamefont
  {Ran}},\ }\href@noop {} {\bibfield  {journal} {\bibinfo  {journal}
  {arXiv:1810.08050}\ } (\bibinfo {year} {2018})}\BibitemShut {NoStop}%
\bibitem [{\citenamefont {Suzuki}\ and\ \citenamefont
  {Inoue}(1987)}]{SI87Trotter}%
  \BibitemOpen
  \bibfield  {author} {\bibinfo {author} {\bibfnamefont {M.}~\bibnamefont
  {Suzuki}}\ and\ \bibinfo {author} {\bibfnamefont {M.}~\bibnamefont {Inoue}},\
  }\href@noop {} {\bibfield  {journal} {\bibinfo  {journal} {Prog. Theor.
  Phys.}\ }\textbf {\bibinfo {volume} {78}},\ \bibinfo {pages} {787} (\bibinfo
  {year} {1987})}\BibitemShut {NoStop}%
\bibitem [{\citenamefont {Inoue}\ and\ \citenamefont
  {Suzuki}(1988)}]{IS88Trotter}%
  \BibitemOpen
  \bibfield  {author} {\bibinfo {author} {\bibfnamefont {M.}~\bibnamefont
  {Inoue}}\ and\ \bibinfo {author} {\bibfnamefont {M.}~\bibnamefont {Suzuki}},\
  }\href@noop {} {\bibfield  {journal} {\bibinfo  {journal} {Progress of
  theoretical physics}\ }\textbf {\bibinfo {volume} {79}},\ \bibinfo {pages}
  {645} (\bibinfo {year} {1988})}\BibitemShut {NoStop}%
\bibitem [{\citenamefont {Bethe}(1935)}]{B1935Bethe}%
  \BibitemOpen
  \bibfield  {author} {\bibinfo {author} {\bibfnamefont {H.~A.}\ \bibnamefont
  {Bethe}},\ }\href@noop {} {\bibfield  {journal} {\bibinfo  {journal} {Proc.
  Roy. Soc. London Ser A.}\ }\textbf {\bibinfo {volume} {150}},\ \bibinfo
  {pages} {552} (\bibinfo {year} {1935})}\BibitemShut {NoStop}%
\bibitem [{\citenamefont {Jiang}\ \emph {et~al.}(2008)\citenamefont {Jiang},
  \citenamefont {Weng},\ and\ \citenamefont {Xiang}}]{JWX08SimpleUpdate}%
  \BibitemOpen
  \bibfield  {author} {\bibinfo {author} {\bibfnamefont {H.-C.}\ \bibnamefont
  {Jiang}}, \bibinfo {author} {\bibfnamefont {Z.-Y.}\ \bibnamefont {Weng}}, \
  and\ \bibinfo {author} {\bibfnamefont {T.}~\bibnamefont {Xiang}},\
  }\href@noop {} {\bibfield  {journal} {\bibinfo  {journal} {Phys. Rev. Lett.}\
  }\textbf {\bibinfo {volume} {101}},\ \bibinfo {pages} {090603} (\bibinfo
  {year} {2008})}\BibitemShut {NoStop}%
\bibitem [{\citenamefont {Ran}(2016)}]{R16AOP}%
  \BibitemOpen
  \bibfield  {author} {\bibinfo {author} {\bibfnamefont {S.-J.}\ \bibnamefont
  {Ran}},\ }\href@noop {} {\bibfield  {journal} {\bibinfo  {journal} {Phys.
  Rev. E}\ }\textbf {\bibinfo {volume} {93}},\ \bibinfo {pages} {053310}
  (\bibinfo {year} {2016})}\BibitemShut {NoStop}%
\bibitem [{\citenamefont {Haldane}(1980)}]{H80Luttinger}%
  \BibitemOpen
  \bibfield  {author} {\bibinfo {author} {\bibfnamefont {F.~D.~M.}\
  \bibnamefont {Haldane}},\ }\href {\doibase 10.1103/PhysRevLett.45.1358}
  {\bibfield  {journal} {\bibinfo  {journal} {Phys. Rev. Lett.}\ }\textbf
  {\bibinfo {volume} {45}},\ \bibinfo {pages} {1358} (\bibinfo {year}
  {1980})}\BibitemShut {NoStop}%
\bibitem [{\citenamefont {Haldane}(1981)}]{H81Luttinger}%
  \BibitemOpen
  \bibfield  {author} {\bibinfo {author} {\bibfnamefont {F.~D.~M.}\
  \bibnamefont {Haldane}},\ }\href
  {http://stacks.iop.org/0022-3719/14/i=19/a=010} {\bibfield  {journal}
  {\bibinfo  {journal} {Journal of Physics C: Solid State Physics}\ }\textbf
  {\bibinfo {volume} {14}},\ \\ \bibinfo {pages} {2585} (\bibinfo {year}
  {1981})}\BibitemShut {NoStop}%
\bibitem [{\citenamefont {Kitaev}(2006)}]{kitaev2006anyons}%
  \BibitemOpen
  \bibfield  {author} {\bibinfo {author} {\bibfnamefont {A.}~\bibnamefont
  {Kitaev}},\ }\href@noop {} {\bibfield  {journal} {\bibinfo  {journal} {Annals
  of Physics}\ }\textbf {\bibinfo {volume} {321}},\ \bibinfo {pages} {2}
  (\bibinfo {year} {2006})}\BibitemShut {NoStop}%
\bibitem [{\citenamefont {Nasu}\ \emph {et~al.}(2015)\citenamefont {Nasu},
  \citenamefont {Udagawa},\ and\ \citenamefont {Motome}}]{NUM15kitaev}%
  \BibitemOpen
  \bibfield  {author} {\bibinfo {author} {\bibfnamefont {J.}~\bibnamefont
  {Nasu}}, \bibinfo {author} {\bibfnamefont {M.}~\bibnamefont {Udagawa}}, \
  and\ \bibinfo {author} {\bibfnamefont {Y.}~\bibnamefont {Motome}},\ }\href
  {\doibase 10.1103/PhysRevB.92.115122} {\bibfield  {journal} {\bibinfo
  {journal} {Phys. Rev. B}\ }\textbf {\bibinfo {volume} {92}},\ \bibinfo
  {pages} {115122} (\bibinfo {year} {2015})}\BibitemShut {NoStop}%
\bibitem [{\citenamefont {Jahromi}\ and\ \citenamefont
  {Orus}(2018)}]{JO18graphPEPS}%
  \BibitemOpen
  \bibfield  {author} {\bibinfo {author} {\bibfnamefont {S.~S.}\ \bibnamefont
  {Jahromi}}\ and\ \bibinfo {author} {\bibfnamefont {R.}~\bibnamefont {Orus}},\
  }\href@noop {} {\bibfield  {journal} {\bibinfo  {journal} {arXiv preprint
  arXiv:1808.00680}\ } (\bibinfo {year} {2018})}\BibitemShut {NoStop}%
\bibitem [{\citenamefont {Sandvik}(1998)}]{S98CubicQMC}%
  \BibitemOpen
  \bibfield  {author} {\bibinfo {author} {\bibfnamefont {A.~W.}\ \bibnamefont
  {Sandvik}},\ }\href {\doibase 10.1103/PhysRevLett.80.5196} {\bibfield
  {journal} {\bibinfo  {journal} {Phys. Rev. Lett.}\ }\textbf {\bibinfo
  {volume} {80}},\ \bibinfo {pages} {5196} (\bibinfo {year}
  {1998})}\BibitemShut {NoStop}%
\bibitem [{\citenamefont {Verstraete}\ \emph {et~al.}(2004)\citenamefont
  {Verstraete}, \citenamefont {Garc\'ia-Ripoll},\ and\ \citenamefont
  {Cirac}}]{VGC04MPDO}%
  \BibitemOpen
  \bibfield  {author} {\bibinfo {author} {\bibfnamefont {F.}~\bibnamefont
  {Verstraete}}, \bibinfo {author} {\bibfnamefont {J.~J.}\ \bibnamefont
  {Garc\'ia-Ripoll}}, \ and\ \bibinfo {author} {\bibfnamefont {J.~I.}\
  \bibnamefont {Cirac}},\ }\href {\doibase 10.1103/PhysRevLett.93.207204}
  {\bibfield  {journal} {\bibinfo  {journal} {Phys. Rev. Lett.}\ }\textbf
  {\bibinfo {volume} {93}},\ \bibinfo {pages} {207204} (\bibinfo {year}
  {2004})}\BibitemShut {NoStop}%
\bibitem [{\citenamefont {Zwolak}\ and\ \citenamefont {Vidal}(2004)}]{ZV04MPO}%
  \BibitemOpen
  \bibfield  {author} {\bibinfo {author} {\bibfnamefont {M.}~\bibnamefont
  {Zwolak}}\ and\ \bibinfo {author} {\bibfnamefont {G.}~\bibnamefont {Vidal}},\
  }\href {\doibase 10.1103/PhysRevLett.93.207205} {\bibfield  {journal}
  {\bibinfo  {journal} {Phys. Rev. Lett.}\ }\textbf {\bibinfo {volume} {93}},\
  \bibinfo {pages} {207205} (\bibinfo {year} {2004})}\BibitemShut {NoStop}%
\bibitem [{\citenamefont {Crosswhite}\ \emph {et~al.}(2008)\citenamefont
  {Crosswhite}, \citenamefont {Doherty},\ and\ \citenamefont
  {Vidal}}]{CDV08MPOLR}%
  \BibitemOpen
  \bibfield  {author} {\bibinfo {author} {\bibfnamefont {G.~M.}\ \bibnamefont
  {Crosswhite}}, \bibinfo {author} {\bibfnamefont {A.~C.}\ \bibnamefont
  {Doherty}}, \ and\ \bibinfo {author} {\bibfnamefont {G.}~\bibnamefont
  {Vidal}},\ }\href {\doibase 10.1103/PhysRevB.78.035116} {\bibfield  {journal}
  {\bibinfo  {journal} {Phys. Rev. B}\ }\textbf {\bibinfo {volume} {78}},\
  \bibinfo {pages} {035116} (\bibinfo {year} {2008})}\BibitemShut {NoStop}%
\bibitem [{\citenamefont {Pirvu}\ \emph {et~al.}(2010)\citenamefont {Pirvu},
  \citenamefont {Murg}, \citenamefont {Cirac},\ and\ \citenamefont
  {Verstraete}}]{PMCV10MPO}%
  \BibitemOpen
  \bibfield  {author} {\bibinfo {author} {\bibfnamefont {B.}~\bibnamefont
  {Pirvu}}, \bibinfo {author} {\bibfnamefont {V.}~\bibnamefont {Murg}},
  \bibinfo {author} {\bibfnamefont {J.~I.}\ \bibnamefont {Cirac}}, \ and\
  \bibinfo {author} {\bibfnamefont {F.}~\bibnamefont {Verstraete}},\ }\href
  {http://stacks.iop.org/1367-2630/12/i=2/a=025012} {\bibfield  {journal}
  {\bibinfo  {journal} {New Journal of Physics}\ }\textbf {\bibinfo {volume}
  {12}},\ \bibinfo {pages} {025012} (\bibinfo {year} {2010})}\BibitemShut
  {NoStop}%
\bibitem [{\citenamefont {Fr\"owis}\ \emph {et~al.}(2010)\citenamefont
  {Fr\"owis}, \citenamefont {Nebendahl},\ and\ \citenamefont
  {D\"ur}}]{FND10MPO2D}%
  \BibitemOpen
  \bibfield  {author} {\bibinfo {author} {\bibfnamefont {F.}~\bibnamefont
  {Fr\"owis}}, \bibinfo {author} {\bibfnamefont {V.}~\bibnamefont {Nebendahl}},
  \ and\ \bibinfo {author} {\bibfnamefont {W.}~\bibnamefont {D\"ur}},\ }\href
  {\doibase 10.1103/PhysRevA.81.062337} {\bibfield  {journal} {\bibinfo
  {journal} {Phys. Rev. A}\ }\textbf {\bibinfo {volume} {81}},\ \bibinfo
  {pages} {062337} (\bibinfo {year} {2010})}\BibitemShut {NoStop}%
\bibitem [{\citenamefont {Becker}\ \emph {et~al.}(2017)\citenamefont {Becker},
  \citenamefont {K\"{o}hler}, \citenamefont {Tiegel}, \citenamefont {Manmana},
  \citenamefont {Wessel},\ and\ \citenamefont {Honecker}}]{BKTMWH17FT1D}%
  \BibitemOpen
  \bibfield  {author} {\bibinfo {author} {\bibfnamefont {J.}~\bibnamefont
  {Becker}}, \bibinfo {author} {\bibfnamefont {T.}~\bibnamefont {K\"{o}hler}},
  \bibinfo {author} {\bibfnamefont {A.~C.}\ \bibnamefont {Tiegel}}, \bibinfo
  {author} {\bibfnamefont {S.~R.}\ \bibnamefont {Manmana}}, \bibinfo {author}
  {\bibfnamefont {S.}~\bibnamefont {Wessel}}, \ and\ \bibinfo {author}
  {\bibfnamefont {A.}~\bibnamefont {Honecker}},\ }\href {\doibase
  10.1103/PhysRevB.96.060403} {\bibfield  {journal} {\bibinfo  {journal} {Phys.
  Rev. B}\ }\textbf {\bibinfo {volume} {96}},\ \bibinfo {pages} {060403}
  (\bibinfo {year} {2017})}\BibitemShut {NoStop}%
\bibitem [{\citenamefont {Gangat}\ \emph {et~al.}(2017)\citenamefont {Gangat},
  \citenamefont {Te},\ and\ \citenamefont {Kao}}]{GIK17FT1D}%
  \BibitemOpen
  \bibfield  {author} {\bibinfo {author} {\bibfnamefont {A.~A.}\ \bibnamefont
  {Gangat}}, \bibinfo {author} {\bibnamefont {Te}}, \ and\ \bibinfo {author}
  {\bibfnamefont {Y.-J.}\ \bibnamefont {Kao}},\ }\href {\doibase
  10.1103/PhysRevLett.119.010501} {\bibfield  {journal} {\bibinfo  {journal}
  {Phys. Rev. Lett.}\ }\textbf {\bibinfo {volume} {119}},\ \bibinfo {pages}
  {010501} (\bibinfo {year} {2017})}\BibitemShut {NoStop}%
\bibitem [{\citenamefont {White}\ and\ \citenamefont
  {Scalapino}(1998)}]{WS98tjDMRG}%
  \BibitemOpen
  \bibfield  {author} {\bibinfo {author} {\bibfnamefont {S.~R.}\ \bibnamefont
  {White}}\ and\ \bibinfo {author} {\bibfnamefont {D.~J.}\ \bibnamefont
  {Scalapino}},\ }\href@noop {} {\bibfield  {journal} {\bibinfo  {journal}
  {Phys. Rev. Lett.}\ }\textbf {\bibinfo {volume} {80}},\ \bibinfo {pages}
  {1272} (\bibinfo {year} {1998})}\BibitemShut {NoStop}%
\bibitem [{\citenamefont {Xiang}\ \emph {et~al.}(2001)\citenamefont {Xiang},
  \citenamefont {Lou},\ and\ \citenamefont {Su}}]{XLS01DMRG2D}%
  \BibitemOpen
  \bibfield  {author} {\bibinfo {author} {\bibfnamefont {T.}~\bibnamefont
  {Xiang}}, \bibinfo {author} {\bibfnamefont {J.}~\bibnamefont {Lou}}, \ and\
  \bibinfo {author} {\bibfnamefont {Z.}~\bibnamefont {Su}},\ }\href {\doibase
  10.1103/PhysRevB.64.104414} {\bibfield  {journal} {\bibinfo  {journal} {Phys.
  Rev. B}\ }\textbf {\bibinfo {volume} {64}},\ \bibinfo {pages} {104414}
  (\bibinfo {year} {2001})}\BibitemShut {NoStop}%
\bibitem [{\citenamefont {Stoudenmire}\ and\ \citenamefont
  {White}(2012)}]{SW12DMRG2DRev}%
  \BibitemOpen
  \bibfield  {author} {\bibinfo {author} {\bibfnamefont {E.~M.}\ \bibnamefont
  {Stoudenmire}}\ and\ \bibinfo {author} {\bibfnamefont {S.~R.}\ \bibnamefont
  {White}},\ }\href@noop {} {\bibfield  {journal} {\bibinfo  {journal} {Ann.
  Rev. Cond. Matter Phys.}\ }\textbf {\bibinfo {volume} {3}},\ \bibinfo {pages}
  {111} (\bibinfo {year} {2012})}\BibitemShut {NoStop}%
\bibitem [{\citenamefont {Prokof'ev}\ \emph {et~al.}(1996)\citenamefont
  {Prokof'ev}, \citenamefont {Svistunov},\ and\ \citenamefont
  {Tupitsyn}}]{PST96QMC}%
  \BibitemOpen
  \bibfield  {author} {\bibinfo {author} {\bibfnamefont {N.~V.}\ \bibnamefont
  {Prokof'ev}}, \bibinfo {author} {\bibfnamefont {B.~V.}\ \bibnamefont
  {Svistunov}}, \ and\ \bibinfo {author} {\bibfnamefont {I.~S.}\ \bibnamefont
  {Tupitsyn}},\ }\href@noop {} {\bibfield  {journal} {\bibinfo  {journal}
  {Journal of Experimental and Theoretical Physics Letters}\ }\textbf {\bibinfo
  {volume} {64}},\ \bibinfo {pages} {911} (\bibinfo {year} {1996})}\BibitemShut
  {NoStop}%
\bibitem [{\citenamefont {Prokof'ev}\ \emph {et~al.}(1998)\citenamefont
  {Prokof'ev}, \citenamefont {Svistunov},\ and\ \citenamefont
  {Tupitsyn}}]{PST98QMC}%
  \BibitemOpen
  \bibfield  {author} {\bibinfo {author} {\bibfnamefont {N.~V.}\ \bibnamefont
  {Prokof'ev}}, \bibinfo {author} {\bibfnamefont {B.~V.}\ \bibnamefont
  {Svistunov}}, \ and\ \bibinfo {author} {\bibfnamefont {I.~S.}\ \bibnamefont
  {Tupitsyn}},\ }\href@noop {} {\bibfield  {journal} {\bibinfo  {journal}
  {Physics Letters A}\ }\textbf {\bibinfo {volume} {238}},\ \bibinfo {pages}
  {253} (\bibinfo {year} {1998})}\BibitemShut {NoStop}%
\bibitem [{\citenamefont {Kashurnikov}\ \emph {et~al.}(1999)\citenamefont
  {Kashurnikov}, \citenamefont {Prokof'ev}, \citenamefont {Svistunov},\ and\
  \citenamefont {Troyer}}]{KPST99QMC}%
  \BibitemOpen
  \bibfield  {author} {\bibinfo {author} {\bibfnamefont {V.~A.}\ \bibnamefont
  {Kashurnikov}}, \bibinfo {author} {\bibfnamefont {N.~V.}\ \bibnamefont
  {Prokof'ev}}, \bibinfo {author} {\bibfnamefont {B.~V.}\ \bibnamefont
  {Svistunov}}, \ and\ \bibinfo {author} {\bibfnamefont {M.}~\bibnamefont
  {Troyer}},\ }\href@noop {} {\bibfield  {journal} {\bibinfo  {journal}
  {Physical Review B}\ }\textbf {\bibinfo {volume} {59}},\ \bibinfo {pages}
  {1162} (\bibinfo {year} {1999})}\BibitemShut {NoStop}%
\bibitem [{\citenamefont {Xi}\ \emph {et~al.}(2011)\citenamefont {Xi},
  \citenamefont {Hu}, \citenamefont {Zhao}, \citenamefont {Su}, \citenamefont
  {Normand},\ and\ \citenamefont {Wang}}]{XHZS+11QMC}%
  \BibitemOpen
  \bibfield  {author} {\bibinfo {author} {\bibfnamefont {B.}~\bibnamefont
  {Xi}}, \bibinfo {author} {\bibfnamefont {S.}~\bibnamefont {Hu}}, \bibinfo
  {author} {\bibfnamefont {J.}~\bibnamefont {Zhao}}, \bibinfo {author}
  {\bibfnamefont {G.}~\bibnamefont {Su}}, \bibinfo {author} {\bibfnamefont
  {B.}~\bibnamefont {Normand}}, \ and\ \bibinfo {author} {\bibfnamefont
  {X.}~\bibnamefont {Wang}},\ }\href@noop {} {\bibfield  {journal} {\bibinfo
  {journal} {Physical Review B}\ }\textbf {\bibinfo {volume} {84}},\ \bibinfo
  {pages} {134407} (\bibinfo {year} {2011})}\BibitemShut {NoStop}%
\bibitem [{\citenamefont {Zhang}\ \emph {et~al.}(2017)\citenamefont {Zhang},
  \citenamefont {Pagano}, \citenamefont {Hess}, \citenamefont {Kyprianidis},
  \citenamefont {Becker}, \citenamefont {Kaplan}, \citenamefont {Gorshkov},
  \citenamefont {Gong},\ and\ \citenamefont {Monroe}}]{zhang2017observation}%
  \BibitemOpen
  \bibfield  {author} {\bibinfo {author} {\bibfnamefont {J.}~\bibnamefont
  {Zhang}}, \bibinfo {author} {\bibfnamefont {G.}~\bibnamefont {Pagano}},
  \bibinfo {author} {\bibfnamefont {P.~W.}\ \bibnamefont {Hess}}, \bibinfo
  {author} {\bibfnamefont {A.}~\bibnamefont {Kyprianidis}}, \bibinfo {author}
  {\bibfnamefont {P.}~\bibnamefont {Becker}}, \bibinfo {author} {\bibfnamefont
  {H.}~\bibnamefont {Kaplan}}, \bibinfo {author} {\bibfnamefont {A.~V.}\
  \bibnamefont {Gorshkov}}, \bibinfo {author} {\bibfnamefont {Z.-X.}\
  \bibnamefont {Gong}}, \ and\ \bibinfo {author} {\bibfnamefont
  {C.}~\bibnamefont {Monroe}},\ }\href@noop {} {\bibfield  {journal} {\bibinfo
  {journal} {Nature}\ }\textbf {\bibinfo {volume} {551}},\ \bibinfo {pages}
  {601} (\bibinfo {year} {2017})}\BibitemShut {NoStop}%
\bibitem [{\citenamefont {Friis}\ \emph {et~al.}(2018)\citenamefont {Friis},
  \citenamefont {Marty}, \citenamefont {Maier}, \citenamefont {Hempel},
  \citenamefont {Holz{\"a}pfel}, \citenamefont {Jurcevic}, \citenamefont
  {Plenio}, \citenamefont {Huber}, \citenamefont {Roos}, \citenamefont {Blatt}
  \emph {et~al.}}]{friis2018observation}%
  \BibitemOpen
  \bibfield  {author} {\bibinfo {author} {\bibfnamefont {N.}~\bibnamefont
  {Friis}}, \bibinfo {author} {\bibfnamefont {O.}~\bibnamefont {Marty}},
  \bibinfo {author} {\bibfnamefont {C.}~\bibnamefont {Maier}}, \bibinfo
  {author} {\bibfnamefont {C.}~\bibnamefont {Hempel}}, \bibinfo {author}
  {\bibfnamefont {M.}~\bibnamefont {Holz{\"a}pfel}}, \bibinfo {author}
  {\bibfnamefont {P.}~\bibnamefont {Jurcevic}}, \bibinfo {author}
  {\bibfnamefont {M.~B.}\ \bibnamefont {Plenio}}, \bibinfo {author}
  {\bibfnamefont {M.}~\bibnamefont {Huber}}, \bibinfo {author} {\bibfnamefont
  {C.}~\bibnamefont {Roos}}, \bibinfo {author} {\bibfnamefont {R.}~\bibnamefont
  {Blatt}},  \emph {et~al.},\ }\href@noop {} {\bibfield  {journal} {\bibinfo
  {journal} {Physical Review X}\ }\textbf {\bibinfo {volume} {8}},\ \bibinfo
  {pages} {021012} (\bibinfo {year} {2018})}\BibitemShut {NoStop}%
\bibitem [{\citenamefont {Bernien}\ \emph {et~al.}(2017)\citenamefont
  {Bernien}, \citenamefont {Schwartz}, \citenamefont {Keesling}, \citenamefont
  {Levine}, \citenamefont {Omran}, \citenamefont {Pichler}, \citenamefont
  {Choi}, \citenamefont {Zibrov}, \citenamefont {Endres}, \citenamefont
  {Greiner} \emph {et~al.}}]{bernien2017probing}%
  \BibitemOpen
  \bibfield  {author} {\bibinfo {author} {\bibfnamefont {H.}~\bibnamefont
  {Bernien}}, \bibinfo {author} {\bibfnamefont {S.}~\bibnamefont {Schwartz}},
  \bibinfo {author} {\bibfnamefont {A.}~\bibnamefont {Keesling}}, \bibinfo
  {author} {\bibfnamefont {H.}~\bibnamefont {Levine}}, \bibinfo {author}
  {\bibfnamefont {A.}~\bibnamefont {Omran}}, \bibinfo {author} {\bibfnamefont
  {H.}~\bibnamefont {Pichler}}, \bibinfo {author} {\bibfnamefont
  {S.}~\bibnamefont {Choi}}, \bibinfo {author} {\bibfnamefont {A.~S.}\
  \bibnamefont {Zibrov}}, \bibinfo {author} {\bibfnamefont {M.}~\bibnamefont
  {Endres}}, \bibinfo {author} {\bibfnamefont {M.}~\bibnamefont {Greiner}},
  \emph {et~al.},\ }\href@noop {} {\bibfield  {journal} {\bibinfo  {journal}
  {Nature}\ }\textbf {\bibinfo {volume} {551}},\ \bibinfo {pages} {579}
  (\bibinfo {year} {2017})}\BibitemShut {NoStop}%
\bibitem [{\citenamefont {Mazurenko}\ \emph {et~al.}(2017)\citenamefont
  {Mazurenko}, \citenamefont {Chiu}, \citenamefont {Ji}, \citenamefont
  {Parsons}, \citenamefont {Kan{\'a}sz-Nagy}, \citenamefont {Schmidt},
  \citenamefont {Grusdt}, \citenamefont {Demler}, \citenamefont {Greif},\ and\
  \citenamefont {Greiner}}]{mazurenko2017cold}%
  \BibitemOpen
  \bibfield  {author} {\bibinfo {author} {\bibfnamefont {A.}~\bibnamefont
  {Mazurenko}}, \bibinfo {author} {\bibfnamefont {C.~S.}\ \bibnamefont {Chiu}},
  \bibinfo {author} {\bibfnamefont {G.}~\bibnamefont {Ji}}, \bibinfo {author}
  {\bibfnamefont {M.~F.}\ \bibnamefont {Parsons}}, \bibinfo {author}
  {\bibfnamefont {M.}~\bibnamefont {Kan{\'a}sz-Nagy}}, \bibinfo {author}
  {\bibfnamefont {R.}~\bibnamefont {Schmidt}}, \bibinfo {author} {\bibfnamefont
  {F.}~\bibnamefont {Grusdt}}, \bibinfo {author} {\bibfnamefont
  {E.}~\bibnamefont {Demler}}, \bibinfo {author} {\bibfnamefont
  {D.}~\bibnamefont {Greif}}, \ and\ \bibinfo {author} {\bibfnamefont
  {M.}~\bibnamefont {Greiner}},\ }\href@noop {} {\bibfield  {journal} {\bibinfo
   {journal} {Nature}\ }\textbf {\bibinfo {volume} {545}},\ \bibinfo {pages}
  {462} (\bibinfo {year} {2017})}\BibitemShut {NoStop}%
\bibitem [{\citenamefont {Rodr\'{\i}guez-Laguna}\ \emph
  {et~al.}(2017)\citenamefont {Rodr\'{\i}guez-Laguna}, \citenamefont
  {Tarruell}, \citenamefont {Lewenstein},\ and\ \citenamefont
  {Celi}}]{RTLC17coldatom}%
  \BibitemOpen
  \bibfield  {author} {\bibinfo {author} {\bibfnamefont {J.}~\bibnamefont
  {Rodr\'{\i}guez-Laguna}}, \bibinfo {author} {\bibfnamefont {L.}~\bibnamefont
  {Tarruell}}, \bibinfo {author} {\bibfnamefont {M.}~\bibnamefont
  {Lewenstein}}, \ and\ \bibinfo {author} {\bibfnamefont {A.}~\bibnamefont
  {Celi}},\ }\href {\doibase 10.1103/PhysRevA.95.013627} {\bibfield  {journal}
  {\bibinfo  {journal} {Phys. Rev. A}\ }\textbf {\bibinfo {volume} {95}},\
  \bibinfo {pages} {013627} (\bibinfo {year} {2017})}\BibitemShut {NoStop}%
\bibitem [{web()}]{websitesqcompute}%
  \BibitemOpen
  \href@noop {} {}\bibinfo {note} {Some websites about quantum computation:
  \url{www.dwavesys.com} (D-waveb);
  \url{ai.google/research/teams/applied-science/quantum-ai} (Google AI);
  \url{www.research.ibm.com/ibm-q} (IBM Q);
  \url{www.microsoft.com/en-us/quantum} (Microsoft)}\BibitemShut {NoStop}%
\bibitem [{\citenamefont {Daley}\ \emph {et~al.}(2004)\citenamefont {Daley},
  \citenamefont {Kollath}, \citenamefont {Schollw\"{o}ck},\ and\ \citenamefont
  {Vidal}}]{DKSV04timeHeff}%
  \BibitemOpen
  \bibfield  {author} {\bibinfo {author} {\bibfnamefont {A.~J.}\ \bibnamefont
  {Daley}}, \bibinfo {author} {\bibfnamefont {C.}~\bibnamefont {Kollath}},
  \bibinfo {author} {\bibfnamefont {U.}~\bibnamefont {Schollw\"{o}ck}}, \ and\
  \bibinfo {author} {\bibfnamefont {G.}~\bibnamefont {Vidal}},\ }\href
  {http://stacks.iop.org/1742-5468/2004/i=04/a=P04005} {\bibfield  {journal}
  {\bibinfo  {journal} {Journal of Statistical Mechanics: Theory and
  Experiment}\ }\textbf {\bibinfo {volume} {2004}} \\ (\bibinfo {year}
  {2004})}\BibitemShut {NoStop}%
\bibitem [{\citenamefont {Phien}\ \emph {et~al.}(2012)\citenamefont {Phien},
  \citenamefont {Vidal},\ and\ \citenamefont {McCulloch}}]{PVM12InfBound}%
  \BibitemOpen
  \bibfield  {author} {\bibinfo {author} {\bibfnamefont {H.~N.}\ \bibnamefont
  {Phien}}, \bibinfo {author} {\bibfnamefont {G.}~\bibnamefont {Vidal}}, \ and\
  \bibinfo {author} {\bibfnamefont {I.~P.}\ \bibnamefont {McCulloch}},\ }\href
  {\doibase 10.1103/PhysRevB.86.245107} {\bibfield  {journal} {\bibinfo
  {journal} {Phys. Rev. B}\ }\textbf {\bibinfo {volume} {86}},\ \bibinfo
  {pages} {245107} (\bibinfo {year} {2012})}\BibitemShut {NoStop}%
\bibitem [{\citenamefont {Chen}\ \emph {et~al.}(2017)\citenamefont {Chen},
  \citenamefont {Liu}, \citenamefont {Chen},\ and\ \citenamefont
  {Li}}]{CLCL17TNexpand}%
  \BibitemOpen
  \bibfield  {author} {\bibinfo {author} {\bibfnamefont {B.-B.}\ \bibnamefont
  {Chen}}, \bibinfo {author} {\bibfnamefont {Y.-J.}\ \bibnamefont {Liu}},
  \bibinfo {author} {\bibfnamefont {Z.}~\bibnamefont {Chen}}, \ and\ \bibinfo
  {author} {\bibfnamefont {W.}~\bibnamefont {Li}},\ }\href {\doibase
  10.1103/PhysRevB.95.161104} {\bibfield  {journal} {\bibinfo  {journal} {Phys.
  Rev. B}\ }\textbf {\bibinfo {volume} {95}},\ \bibinfo {pages} {161104}
  (\bibinfo {year} {2017})}\BibitemShut {NoStop}%
\bibitem [{\citenamefont {Lubasch}\ \emph {et~al.}(2014)\citenamefont
  {Lubasch}, \citenamefont {Cirac},\ and\ \citenamefont
  {Ba\~{n}uls}}]{LCB14FinitePEPS}%
  \BibitemOpen
  \bibfield  {author} {\bibinfo {author} {\bibfnamefont {M.}~\bibnamefont
  {Lubasch}}, \bibinfo {author} {\bibfnamefont {J.~I.}\ \bibnamefont {Cirac}},
  \ and\ \bibinfo {author} {\bibfnamefont {M.-C.}\ \bibnamefont {Ba\~{n}uls}},\
  }\href@noop {} {\bibfield  {journal} {\bibinfo  {journal} {Phys. Rev. B}\
  }\textbf {\bibinfo {volume} {90}},\ \bibinfo {pages} {064425} (\bibinfo
  {year} {2014})}\BibitemShut {NoStop}%
\bibitem [{\citenamefont {Vidal}(2007)}]{V07iTEBD}%
  \BibitemOpen
  \bibfield  {author} {\bibinfo {author} {\bibfnamefont {G.}~\bibnamefont
  {Vidal}},\ }\href@noop {} {\bibfield  {journal} {\bibinfo  {journal} {Phys.
  Rev. Lett.}\ }\textbf {\bibinfo {volume} {98}},\ \bibinfo {pages} {070201}
  (\bibinfo {year} {2007})}\BibitemShut {NoStop}%
\bibitem [{\citenamefont {Jordan}\ \emph {et~al.}(2008)\citenamefont {Jordan},
  \citenamefont {Or\'{u}s}, \citenamefont {Vidal}, \citenamefont {Verstraete},\
  and\ \citenamefont {Cirac}}]{JOVVC08PEPS}%
  \BibitemOpen
  \bibfield  {author} {\bibinfo {author} {\bibfnamefont {J.}~\bibnamefont
  {Jordan}}, \bibinfo {author} {\bibfnamefont {R.}~\bibnamefont {Or\'{u}s}},
  \bibinfo {author} {\bibfnamefont {G.}~\bibnamefont {Vidal}}, \bibinfo
  {author} {\bibfnamefont {F.}~\bibnamefont {Verstraete}}, \ and\ \bibinfo
  {author} {\bibfnamefont {J.~I.}\ \bibnamefont {Cirac}},\ }\href@noop {}
  {\bibfield  {journal} {\bibinfo  {journal} {Phys. Rev. Lett.}\ }\textbf
  {\bibinfo {volume} {101}},\ \bibinfo {pages} {250602} (\bibinfo {year}
  {2008})}\BibitemShut {NoStop}%
\bibitem [{\citenamefont {Nishino}\ and\ \citenamefont
  {Okunishi}(1996)}]{NO96CTMRG0}%
  \BibitemOpen
  \bibfield  {author} {\bibinfo {author} {\bibfnamefont {T.}~\bibnamefont
  {Nishino}}\ and\ \bibinfo {author} {\bibfnamefont {K.}~\bibnamefont
  {Okunishi}},\ }\href@noop {} {\bibfield  {journal} {\bibinfo  {journal} {J.
  Phys. Soc. Jpn.}\ }\textbf {\bibinfo {volume} {65}},\ \bibinfo {pages} {891}
  (\bibinfo {year} {1996})}\BibitemShut {NoStop}%
\bibitem [{\citenamefont {Anders}\ \emph {et~al.}(2010)\citenamefont {Anders},
  \citenamefont {Gull}, \citenamefont {Pollet}, \citenamefont {Troyer},\ and\
  \citenamefont {Werner}}]{AGPTW10DMF}%
  \BibitemOpen
  \bibfield  {author} {\bibinfo {author} {\bibfnamefont {P.}~\bibnamefont
  {Anders}}, \bibinfo {author} {\bibfnamefont {E.}~\bibnamefont {Gull}},
  \bibinfo {author} {\bibfnamefont {L.}~\bibnamefont {Pollet}}, \bibinfo
  {author} {\bibfnamefont {M.}~\bibnamefont {Troyer}}, \ and\ \bibinfo {author}
  {\bibfnamefont {P.}~\bibnamefont {Werner}},\ }\href@noop {} {\bibfield
  {journal} {\bibinfo  {journal} {Phys. Rev. Lett.}\ }\textbf {\bibinfo
  {volume} {105}},\ \bibinfo {pages} {096402} (\bibinfo {year}
  {2010})}\BibitemShut {NoStop}%
\bibitem [{\citenamefont {Georges}\ \emph {et~al.}(1996)\citenamefont
  {Georges}, \citenamefont {Kotliar}, \citenamefont {Krauth},\ and\
  \citenamefont {Rozenberg}}]{GKKR96DMFTRev}%
  \BibitemOpen
  \bibfield  {author} {\bibinfo {author} {\bibfnamefont {A.}~\bibnamefont
  {Georges}}, \bibinfo {author} {\bibfnamefont {G.}~\bibnamefont {Kotliar}},
  \bibinfo {author} {\bibfnamefont {W.}~\bibnamefont {Krauth}}, \ and\ \bibinfo
  {author} {\bibfnamefont {M.~J.}\ \bibnamefont {Rozenberg}},\ }\href@noop {}
  {\bibfield  {journal} {\bibinfo  {journal} {Rev. Mod. Phys.}\ }\textbf
  {\bibinfo {volume} {13}},\ \bibinfo {pages} {68} (\bibinfo {year}
  {1996})}\BibitemShut {NoStop}%
\bibitem [{\citenamefont {Kotliar}\ \emph {et~al.}(2006)\citenamefont
  {Kotliar}, \citenamefont {Savrasov}, \citenamefont {Haule}, \citenamefont
  {Oudovenko}, \citenamefont {Parcollet},\ and\ \citenamefont
  {Marianetti}}]{KSHO+06DMFTrev}%
  \BibitemOpen
  \bibfield  {author} {\bibinfo {author} {\bibfnamefont {G.}~\bibnamefont
  {Kotliar}}, \bibinfo {author} {\bibfnamefont {S.~Y.}\ \bibnamefont
  {Savrasov}}, \bibinfo {author} {\bibfnamefont {K.}~\bibnamefont {Haule}},
  \bibinfo {author} {\bibfnamefont {V.~S.}\ \bibnamefont {Oudovenko}}, \bibinfo
  {author} {\bibfnamefont {O.}~\bibnamefont {Parcollet}}, \ and\ \bibinfo
  {author} {\bibfnamefont {C.~A.}\ \bibnamefont {Marianetti}},\ }\href
  {\doibase 10.1103/RevModPhys.78.865} {\bibfield  {journal} {\bibinfo
  {journal} {Rev. Mod. Phys.}\ }\textbf {\bibinfo {volume} {78}},\ \bibinfo
  {pages} {865} (\bibinfo {year} {2006})}\BibitemShut {NoStop}%
\bibitem [{\citenamefont {Knizia}\ and\ \citenamefont {Chan}(2012)}]{KC12DMET}%
  \BibitemOpen
  \bibfield  {author} {\bibinfo {author} {\bibfnamefont {G.}~\bibnamefont
  {Knizia}}\ and\ \bibinfo {author} {\bibfnamefont {G.~K.~L.}\ \bibnamefont
  {Chan}},\ }\href@noop {} {\bibfield  {journal} {\bibinfo  {journal} {Phys.
  Rev. Lett.}\ }\textbf {\bibinfo {volume} {109}},\ \bibinfo {pages} {186404}
  (\bibinfo {year} {2012})}\BibitemShut {NoStop}%
\bibitem [{\citenamefont {Knizia}\ and\ \citenamefont {Chan}(2013)}]{KC13DMET}%
  \BibitemOpen
  \bibfield  {author} {\bibinfo {author} {\bibfnamefont {G.}~\bibnamefont
  {Knizia}}\ and\ \bibinfo {author} {\bibfnamefont {G.~K.~L.}\ \bibnamefont
  {Chan}},\ }\href@noop {} {\bibfield  {journal} {\bibinfo  {journal} {J. Chem.
  Theor. Comp.}\ }\textbf {\bibinfo {volume} {9}},\ \bibinfo {pages} {1428}
  (\bibinfo {year} {2013})}\BibitemShut {NoStop}%
\bibitem [{\citenamefont {Verstraete}\ and\ \citenamefont
  {Cirac}(2010)}]{VC10cMPS}%
  \BibitemOpen
  \bibfield  {author} {\bibinfo {author} {\bibfnamefont {F.}~\bibnamefont
  {Verstraete}}\ and\ \bibinfo {author} {\bibfnamefont {J.~I.}\ \bibnamefont
  {Cirac}},\ }\href@noop {} {\bibfield  {journal} {\bibinfo  {journal} {Phys.
  Rev. Lett.}\ }\textbf {\bibinfo {volume} {104}},\ \bibinfo {pages} {190405}
  (\bibinfo {year} {2010})}\BibitemShut {NoStop}%
\bibitem [{\citenamefont {Jennings}\ \emph {et~al.}(2015)\citenamefont
  {Jennings}, \citenamefont {Brockt}, \citenamefont {Haegeman}, \citenamefont
  {Osborne},\ and\ \citenamefont {Verstraete}}]{JBHOV15cTNS}%
  \BibitemOpen
  \bibfield  {author} {\bibinfo {author} {\bibfnamefont {D.}~\bibnamefont
  {Jennings}}, \bibinfo {author} {\bibfnamefont {C.}~\bibnamefont {Brockt}},
  \bibinfo {author} {\bibfnamefont {J.}~\bibnamefont {Haegeman}}, \bibinfo
  {author} {\bibfnamefont {T.~J.}\ \bibnamefont {Osborne}}, \ and\ \bibinfo
  {author} {\bibfnamefont {F.}~\bibnamefont {Verstraete}},\ }\href {\doibase
  10.1088/1367-2630/17/6/063039} {\bibfield  {journal} {\bibinfo  {journal}
  {New Journal of Physics}\ }\textbf {\bibinfo {volume} {17}},\ \bibinfo
  {pages} {063039} (\bibinfo {year} {2015})}\BibitemShut {NoStop}%
\bibitem [{\citenamefont {Garc\'ia}\ \emph {et~al.}(2004)\citenamefont
  {Garc\'ia}, \citenamefont {Hallberg},\ and\ \citenamefont
  {Rozenberg}}]{GHR04DMFTwithDMRG}%
  \BibitemOpen
  \bibfield  {author} {\bibinfo {author} {\bibfnamefont {D.~J.}\ \bibnamefont
  {Garc\'ia}}, \bibinfo {author} {\bibfnamefont {K.}~\bibnamefont {Hallberg}},
  \ and\ \bibinfo {author} {\bibfnamefont {M.~J.}\ \bibnamefont {Rozenberg}},\
  }\href@noop {} {\bibfield  {journal} {\bibinfo  {journal} {Phys. Rev. Lett.}\
  }\textbf {\bibinfo {volume} {93}},\ \bibinfo {pages} {246403} (\bibinfo
  {year} {2004})}\BibitemShut {NoStop}%
\bibitem [{\citenamefont {Bauernfeind}\ \emph {et~al.}(2017)\citenamefont
  {Bauernfeind}, \citenamefont {Zingl}, \citenamefont {Triebl}, \citenamefont
  {Aichhorn},\ and\ \citenamefont {Evertz}}]{BZTAE17TNDMFT}%
  \BibitemOpen
  \bibfield  {author} {\bibinfo {author} {\bibfnamefont {D.}~\bibnamefont
  {Bauernfeind}}, \bibinfo {author} {\bibfnamefont {M.}~\bibnamefont {Zingl}},
  \bibinfo {author} {\bibfnamefont {R.}~\bibnamefont {Triebl}}, \bibinfo
  {author} {\bibfnamefont {M.}~\bibnamefont {Aichhorn}}, \ and\ \bibinfo
  {author} {\bibfnamefont {H.~G.}\ \bibnamefont {Evertz}},\ }\href {\doibase
  10.1103/PhysRevX.7.031013} {\bibfield  {journal} {\bibinfo  {journal} {Phys.
  Rev. X}\ }\textbf {\bibinfo {volume} {7}},\ \bibinfo {pages} {031013}
  (\bibinfo {year} {2017})}\BibitemShut {NoStop}%
\end{thebibliography}
%


\end{document}